\documentclass[twocolumn,twoside]{IEEEtran}                 

\usepackage{subfigure}
\usepackage{bm}
\usepackage{graphicx}
\usepackage{amsmath, amsthm, amsfonts, amssymb, amsbsy}
\usepackage{setspace}
\usepackage{graphicx}
\usepackage{pstricks,arydshln}
\usepackage{multirow}
\usepackage{booktabs}
\usepackage{stfloats}
\usepackage{caption}
\usepackage{cite}
\usepackage{epstopdf}
\usepackage{dsfont}
\usepackage{color}
\usepackage{graphicx}
\usepackage{epstopdf}
\usepackage{amsmath, amsthm, amsfonts, amssymb, amsbsy}
\usepackage{setspace}
\usepackage{pstricks,arydshln}
\usepackage{multirow}
\usepackage{booktabs}
\usepackage{stfloats}
\usepackage{subfigure}
\usepackage{caption}
\usepackage{cite}
\usepackage{xcolor}
\usepackage{epstopdf}
\usepackage{cases}
\usepackage[noend]{algpseudocode}
\usepackage{algorithmicx,algorithm}
\usepackage{flushend,cuted}

\newcommand{\RNum}[1]{\uppercase\expandafter{\romannumeral #1\relax}}
\begin{document}
\title{Joint Spatial Division and Coaxial Multiplexing for Downlink Multi-User OAM Wireless Backhaul}

\author{Wen-Xuan Long,~\IEEEmembership{Graduate Student Member,~IEEE,} Rui Chen,~\IEEEmembership{Member,~IEEE,} \\
Marco Moretti,~\IEEEmembership{Member,~IEEE,} Jian Xiong,~\IEEEmembership{Member,~IEEE,} and Jiandong Li,~\IEEEmembership{Fellow,~IEEE}

\thanks{This work was supported in part by Natural Science Basic Research Program of Shaanxi (Program No. 2021JZ-18), Natural Science Foundation of Guangdong Province of China under Grant 2021A1515010812, the open research fund of National Mobile Communications Research Laboratory, Southeast University under Grant number 2021D04.}
\thanks{W.-X. Long is with the State Key Laboratory of ISN, Xidian University, Shaanxi 710071, China, and also with the University of Pisa, Dipartimento di Ingegneria dell'Informazione, Italy (e-mail: wxlong@stu.xidian.edu.cn).}
\thanks{R. Chen is with the State Key Laboratory of ISN, Xidian University, Xi'an 710071, China, and also with the National Mobile Communications Research Laboratory, Southeast University, Nanjing 210018, China (e-mail: rchen@xidian.edu.cn).}
\thanks{M. Moretti is with the University of Pisa, Dipartimento di Ingegneria dell'Informazione, Italy (e-mail: marco.moretti@iet.unipi.it).}
\thanks{J. Xiong is with the Department of Electronic Engineering, Shanghai Jiao Tong University, Shanghai 200240, China (e-mail: xjarrow@sjtu.edu.cn).}
\thanks{J. Li is with the State Key Laboratory of Integrated Service Networks (ISN), Xidian University, Shaanxi 710071, China (e-mail: jdli@mail.xidian.edu.cn).}
}

\maketitle

\thispagestyle{empty}
\begin{abstract}
Orbital angular momentum (OAM) at radio frequency (RF) provides a novel approach of multiplexing a set of orthogonal modes on the same frequency channel to achieve high spectral efficiencies (SEs). However, the existing research on OAM wireless communications is mainly focused on point-to-point transmission in the line-of-sight (LoS) scenario. In this paper, we propose an overall scheme of the downlink multi-user OAM (MU-OAM) wireless backhaul based on uniform circular arrays (UCAs) for broadcasting networks, which can achieve the joint spatial division and coaxial multiplexing (JSDCM). A salient feature of the proposed downlink MU-OAM wireless backhaul systems is that the channel matrices are completely characterized by the position of each small base station (SBS), independent of the numbers of subcarriers and antennas, which avoids estimating large channel matrices required by the traditional downlink multi-user multiple-input multiple-output (MU-MIMO) wireless backhaul systems. Thereafter, we propose an OAM-based multi-user distance and angle of arrival (AoA) estimation method, which is able to simultaneously estimate the positions of multiple SBSs with a flexible number of training symbols. With the estimated distances and AoAs, a MU-OAM preprocessing scheme is applied to eliminate the co-mode and inter-mode interferences in the downlink MU-OAM channel. At last, the proposed methods are extended to the downlink MU-OAM-MIMO wireless backhaul system equipped with uniform concentric circular arrays (UCCAs), for which much higher spectral efficiency (SE) and energy efficiency (EE) than traditional MU-MIMO systems can be achieved. Both mathematical analysis and simulation results validate that the proposed scheme can effectively eliminate both interferences of the practical downlink MU-OAM channel and approaches the performance of the ideal MU-OAM channel.
\end{abstract}


\vspace{0.0cm}
\section{Introduction}
\vspace{0.0cm}

\begin{figure}[t] 
\setlength{\abovecaptionskip}{0cm}   
\setlength{\belowcaptionskip}{-0.2cm}   
\begin{center}
\includegraphics[width=8.0cm,height=4.7cm]{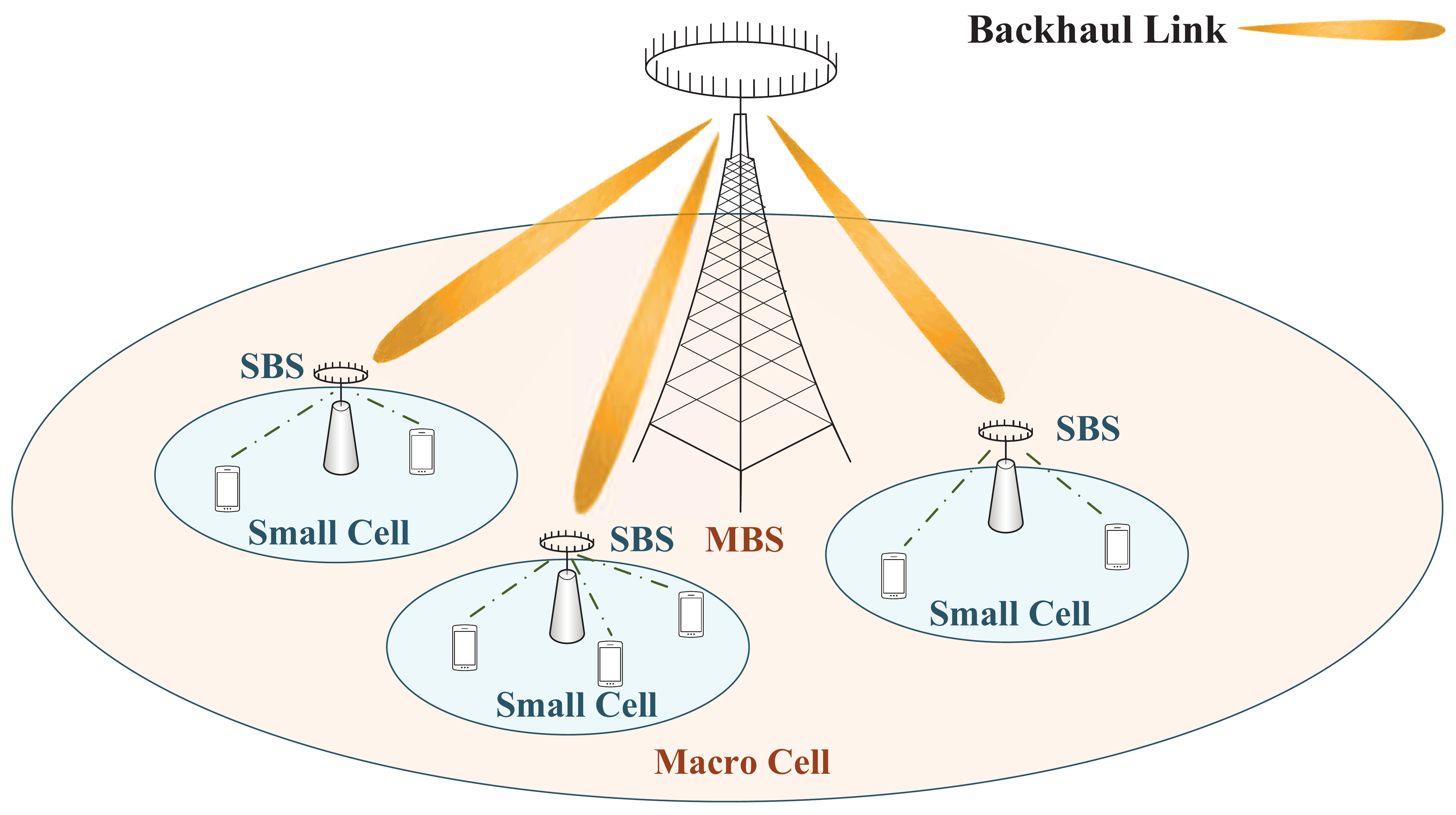}
\end{center}
\caption{The UCA-based downlink MU-OAM wireless backhaul system.}
\label{Fig1}
\end{figure}

The exponential growth of demand for mobile multimedia services is becoming the driving force behind the evolution of wireless communication systems. The rapid development of emerging applications, such as high-definition (HD) video, virtual reality (VR) and auto-pilot driving, results in a never-ending growth in mobile data traffic. It is reported in \cite{Zhang20196} that the next 6G communication system will be required to increase the capacity of current 5G systems hundred fold. To meet the requirement, more and more high frequency bands such as millimeter wave and terahertz bands are being licensed \cite{WRC}. Since radio frequency (RF) spectrum resources are scarce, besides exploiting more frequency bandwidth, innovative frequency reuse techniques are becoming increasingly important for communication systems. The multi-tier heterogeneous network (HetNet), incorporated with flexible backhaul connections, distributed caching and multiple-input multiple-output (MIMO) technologies \cite{Hossain2014Evolution,Xiong2020Distributed,WU2018Nonbinary,Shitomi2018MIMO,Baek2019Implementation,Garro2020Layered}, is a promising solution for next-generation broadcasting networks including digital video broadcasting-next generation handheld (DVB-NGH) \cite{DVB2014Barquero} and advanced television systems committee (ATSC) 3.0 \cite{Fay2016An}. Typically, HetNet consists of a macro base station (MBS) tier and a small base station (SBS) tier as shown in Fig.\ref{Fig1}. Backhaul links between MBS and SBSs exploit wired and wireless connectivity. In 6G systems, considering the massive number of small cells and the implementation costs, wireless backhaul \cite{Wang2015Backhauling,Siddique2015Wireless} will become a mainstream approach to handle the backhaul connectivity of small cells.

\begin{figure}[tb]
\setlength{\abovecaptionskip}{0.1cm}   
\setlength{\belowcaptionskip}{-0.2cm}   
\centering
\subfigure[]{
\includegraphics[width=8.0cm,height=2.0cm]{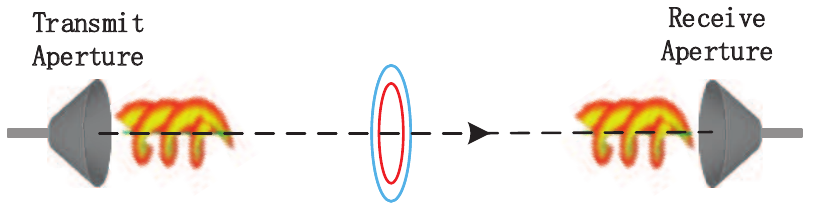}
}

\subfigure[]{
\includegraphics[width=8.0cm,height=5.2cm]{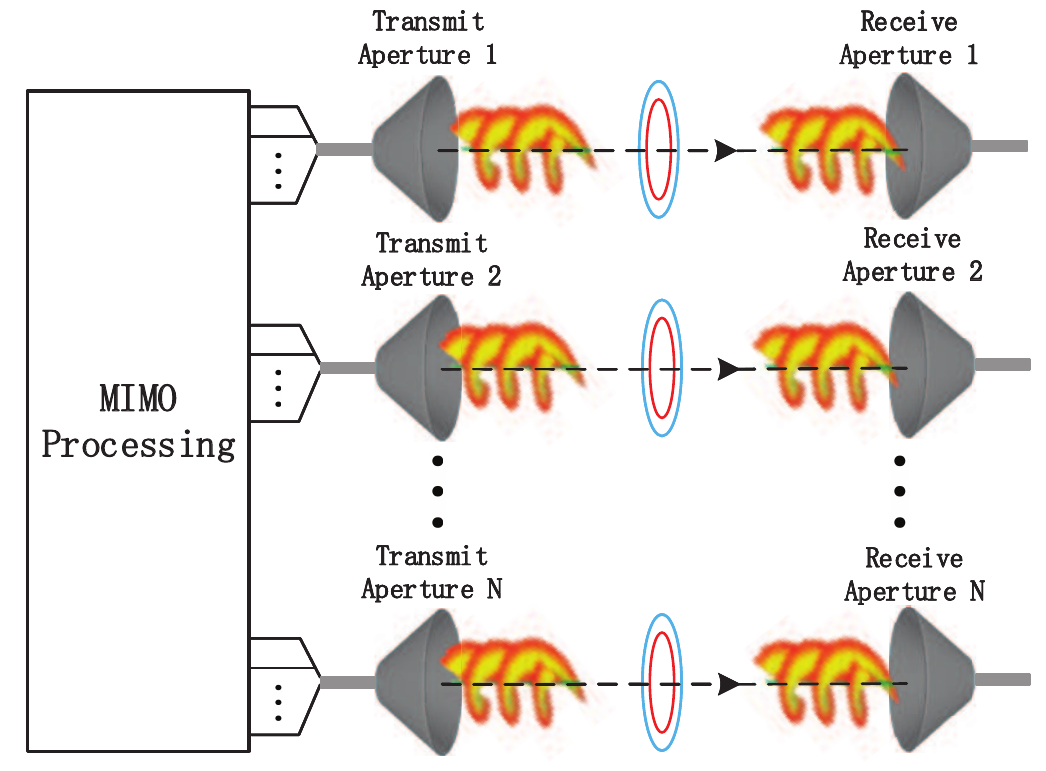}
}

\subfigure[]{
\includegraphics[width=8.0cm,height=5.5cm]{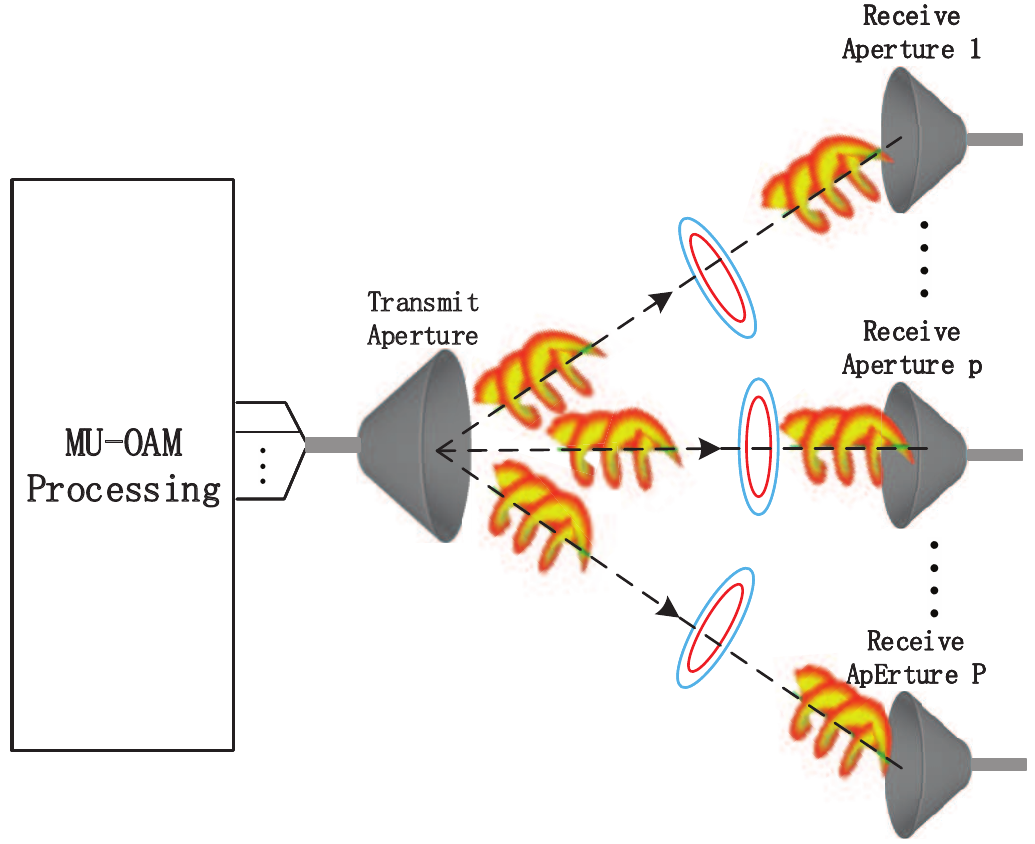}
}
\caption{(a) The coaxial multiplexing of OAM wireless communications, (b) The OAM and MIMO combined communication scheme in \cite{Ren2017Line}, (c) The proposed joint spatial division and coaxial multiplexing scheme for the downlink MU-OAM wireless backhaul.}
\label{Fig2}
\end{figure}

One potential approach for 6G wireless backhaul is the orbital angular momentum (OAM)-based wireless backhaul scheme. The phase front of an electromagnetic (EM) wave carrying OAM rotates with azimuth exhibiting a helical structure $e^{j\ell\phi'}$ in space, where $\phi'$ is the transverse azimuth and $\ell$ is an unbounded integer defined as OAM \emph{topological charge} or OAM \emph{mode number} \cite{Allen1992Orbital}. Due to inherent orthogonality among different OAM modes, the OAM-based wireless communication enables a novel coaxial multiplexing approach as shown in Fig.\ref{Fig2} (a), which utilizes a set of information-bearing modes on the same frequency channel to achieve a high spectral efficiency (SE) \cite{Chen2019Orbital,Tamburini2012Encoding,Mahmouli20134,Yan2014High,Zhang2017Mode,Ren2017Line, Chen2018A,Zhang2019Orbital,Zhao2019Compound,Chen2020Multi}.

%
%

Up to now, a significant progress in the research of OAM-based wireless communications has been made \cite{Chen2019Orbital}. In \cite{Tamburini2012Encoding}, the OAM-based wireless transmission experiment is realized for the first time, which successfully multiplexes two radio signals at the same frequency. Moreover, a 4 Gbps uncompressed video transmission link over a 60 GHz OAM radio channel is implemented in \cite{Mahmouli20134}. It is shown in \cite{Yan2014High} that multiplexing 4 dual-polarized OAM modes (mode number $\ell=\pm1,\pm3$) generated by SPPs can achieve 32 Gbit/s data rate in a wireless communication link at 28 GHz. Besides, it is shown in \cite{Zhang2017Mode} that since the inter-channel interference can be mitigated by the inherent orthogonality between OAM modes, OAM receivers have lower complexity than traditional MIMO receivers. In \cite{Ren2017Line}, a $2\times2$ antenna aperture architecture, where each aperture multiplexes two OAM modes, is implemented in the 28 GHz band achieving a 16 Gbit/s transmission rate. In \cite{Chen2018A}, the transceiver architecture for broadband OAM orthogonal frequency division multiplexing (OAM-OFDM) wireless communication systems is proposed. In \cite{Chen2020Multi}, a complete OAM-based point-to-point wireless communication scheme is proposed in the line-of-sight (LoS) scenario.

However, the existing research on OAM wireless communications is mainly focused on the point-to-point coaxial transmission in the LoS scenario \cite{Tamburini2012Encoding,Mahmouli20134,Yan2014High,Zhang2017Mode,Chen2018A,Yan201632,Chen2018Beam,Zhao2019Compound,Chen2020Multi}, i.e., the single-user OAM (SU-OAM) communications, few paper considers the multi-user OAM (MU-OAM) communications. To the best of our knowledge, only an OAM combined with a MIMO multi-user communication scheme is proposed in \cite{Ren2017Line}, shown in Fig.\ref{Fig2} (b), which lays the foundation for the research of MU-OAM wireless communications. Inspired by the work in \cite{Ren2017Line}, we consider a downlink MU-OAM wireless communication scheme for HetNet wireless backhaul systems with the positions of MBS and each SBS as shown in Fig.\ref{Fig1}, which is illustrated in Fig.\ref{Fig2} (c) for comparison with the existing two OAM multiplexing schemes. Furthermore, we propose an overall communication scheme for the downlink MU-OAM and MU-OAM-MIMO wireless backhaul system including the OAM-based multi-user distance and angle of arrival (AoA) estimation and the preprocessing of MU-OAM signals. The novelty and major contributions of this paper are summarized as follows:
\begin{itemize}
\item
First, we propose a downlink MU-OAM wireless backhaul system based on uniform circular arrays (UCAs), which enables joint spatial division and coaxial multiplexing (JSDCM). Compared with the space division multiplexing in traditional downlink MU-MIMO wireless backhaul system, JSDCM in MU-OAM system can simultaneously multiplex the axis space between the transmit UCA center and each receive UCA center and the radial space perpendicular to the axis space due to the inherent orthogonality among different OAM modes. Therefore, in contrast to the traditional multi-user MIMO (MU-MIMO) system, the proposed MU-OAM system is expected to provide a higher SE.
\item
Second, we formulate the signal models of training and data transmission stages for the proposed MU-OAM system. It's worth noting that the channel matrix of the proposed system is completely characterized by the position of the UCA center in each SBS. According to this character, we propose an effective OAM-based multi-user distance and AoA estimation method to simultaneously estimate the positions of all the SBSs during the training stage. Compared with the traditional MU-MIMO system, the proposed MU-OAM system does not need to estimate a large channel matrix, which can greatly reduce the required training overhead.
\item
Third, we apply a MU-OAM preprocessing scheme to eliminate the co-mode interference and the inter-mode interference in the downlink MU-OAM channel. Since interferences in the system are eliminated through preprocessing at MBS, each SBS only needs to separate and despiralize its own received OAM signals, which reduces the complexity of SBSs. Furthermore, we extend the proposed methods to the downlink MU-OAM-MIMO wireless backhaul system based on uniform concentric circular arrays (UCCAs), for which much higher SE and energy efficiency (EE) than traditional MU-MIMO systems could be achieved.
\end{itemize}

The remainder of this paper is organized as follows. Based on the feasibility of generating and receiving multi-mode OAM beams by UCA \cite{Mohammadi2010system,Liu2016Generation,Chen2018A}, we model the UCA-based downlink MU-OAM wireless backhaul system in Section II. In Section III, the OAM-based multi-user distance and AoA estimation method is proposed. With the estimated distances and AoAs, a MU-OAM preprocessing scheme is applied to effectively eliminate co-mode and inter-mode interferences in the downlink MU-OAM channel in Section IV. In Section V, the proposed methods are extended to the UCCA-based downlink MU-OAM-MIMO wireless backhaul system to achieve higher SE. Simulation results are shown in Section VI and conclusions are summarized in Section VII.

\begin{figure}[t] 
\setlength{\abovecaptionskip}{0cm}   
\setlength{\belowcaptionskip}{-0.2cm}   
\begin{center}
\includegraphics[width=8.2cm,height=6.2cm]{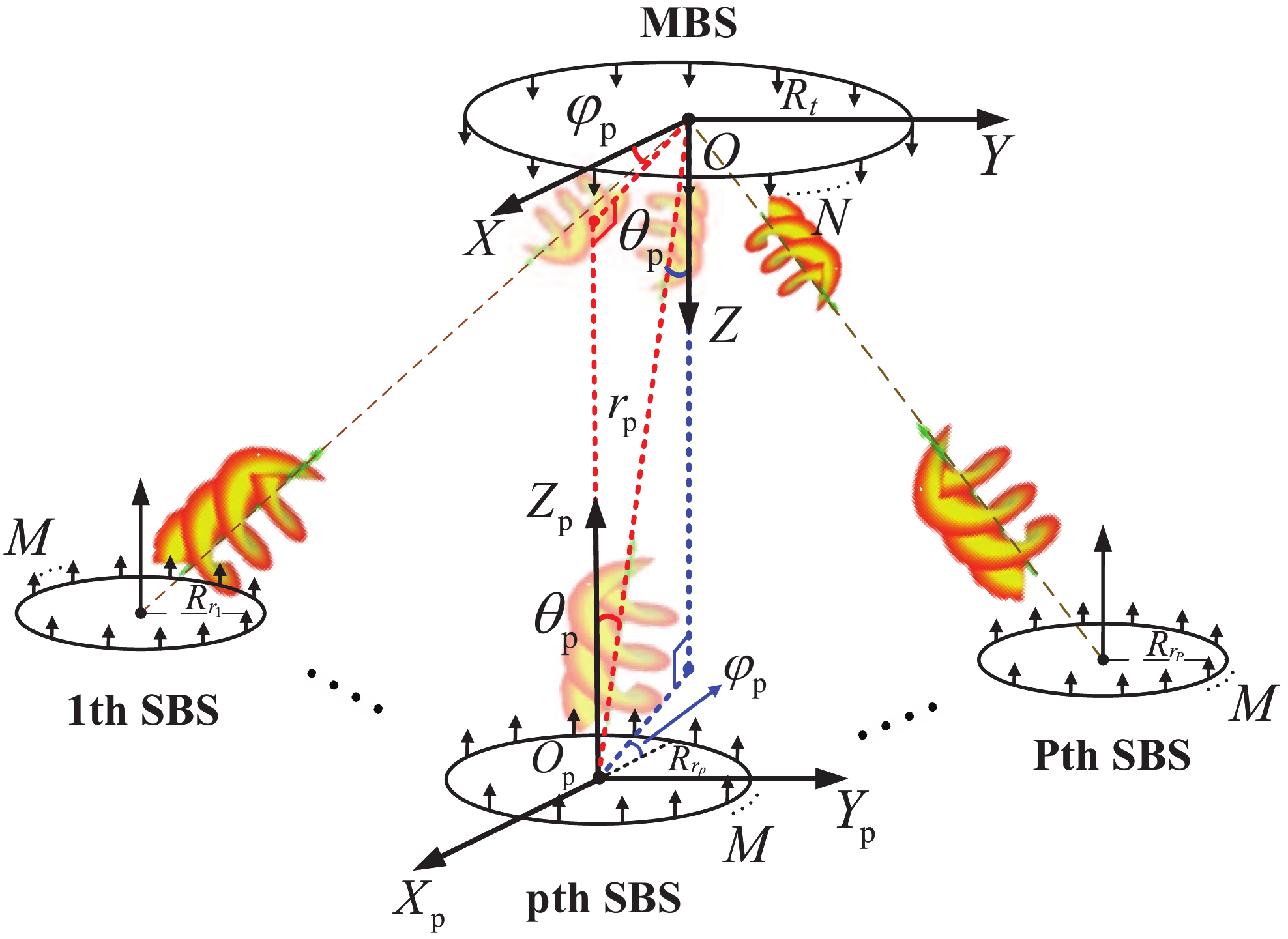}
\end{center}
\caption{The geometrical model of the transmit and receive UCAs in the downlink MU-OAM wireless backhaul system.}
\label{Fig3}
\end{figure}

\vspace{0.0cm}
\section{UCA-Based Downlink MU-OAM Wireless Backhaul System}
\vspace{0.0cm}

Employing UCA is a popular way to generate and receive radio OAM beams due to simple structure and multi-mode multiplexing ability \cite{Liu2016Generation, Chen2018A}. Thus, we consider a UCA-based downlink MU-OAM wireless backhaul system, where multi-mode OAM beams are generated by an $N$-element UCA at MBS and separately received by $P$ $M$-element UCAs at SBSs.

In the downlink MU-OAM wireless backhaul system, the UCA at MBS and the UCAs at $P$ SBSs are off-axis misaligned as shown in Fig.\ref{Fig3} \cite{Chen2018Beam}, where the UCA at MBS and the UCAs at $P$ SBSs are parallel, and the UCAs at $P$ SBSs are not centered with the same axis. Moreover, for the $M$-element UCAs at SBSs, at most $M$ OAM modes can be resolved, i.e., $|\ell|$ $<$ $ M/2$ \cite{Mohammadi2010system}. Hence, the $N$-element UCA at MBS is required to generate $P$ $M$-mode OAM beams for $P$ SBSs in the proposed MU-OAM system. Owing to the limitation on the number of multiplexed OAM modes, $N\geq PM$. Without loss of generality, we assume $N = PM$ here.

\vspace{-0.2cm}
\subsection{Downlink Channel Model}%
\vspace{0.0cm}
In free space communications, propagation through the RF channel leads to attenuation and phase rotation of the transmitted signal. This effect is modeled through multiplying by a complex constant $h$, whose value depends on the frequency and the distance $d$ between the transmit and receive antennas \cite{Edfors2012Is}:
\begin{equation} \label{FreeSpaceChannel}
h(k,d)=\frac{\beta}{2kd}\textrm{exp}\left(-ikd\right),
\end{equation}
where $i=\sqrt{-1}$ is the imaginary unit, $k=2\pi/\lambda$ is the wave number, $\lambda$ is the wavelength, $\beta$ models all constants relative to the antenna elements and their patterns, $1/(2kd)$ denotes the degradation of amplitude, and the complex exponential term is the phase difference due to the propagation distance.

The geometrical model of the UCAs at MBS and the $p$th SBS in the off-axis case is illustrated in Fig.\ref{Fig3}. In the geometrical model, the coordinate system $\textrm{Z}-\textrm{X}\textrm{O}\textrm{Y}$ is established at MBS using the MBS UCA plane as the $\textrm{X}\textrm{O}\textrm{Y}$ plane and the axis through the point $\textrm{O}$ and perpendicular to the UCA plane as the $\textrm{Z}$-axis, and the coordinate system $\textrm{Z}_{p}-\textrm{X}_{p}\textrm{O}_{p}\textrm{Y}_{p}$ is established at the $p$th SBS by the similar approach. According to the geometric relationship in Fig.\ref{Fig3}, the coordinate of the UCA center at the $p$th SBS is denoted as $\textrm{O}_{p}(r_{p},\theta_p,\varphi_p)$ in $\textrm{Z}-\textrm{X}\textrm{O}\textrm{Y}$, and the coordinate of the UCA center at MBS can be denoted as $\textrm{O}(r_{p},\theta_p,\varphi_p +\pi)$ in $\textrm{Z}_p-\textrm{X}_p\textrm{O}_p\textrm{Y}_p$, where $r_p$ is the distance between the UCA center of MBS and the UCA center of the $p$th SBS, $\varphi_p$ and $\theta_p$ are respectively the azimuth angle and the elevation angle of the UCA center at the $p$th SBS in $\textrm{Z}-\textrm{X}\textrm{O}\textrm{Y}$, and \emph{$\theta_p$ and $\varphi_p$ are defined as the angle of arrival (AoA) of OAM beams transmitted by the $p$th SBS.}

\begin{figure*}[t]
\centering
\begin{minipage}{5.3cm}
\includegraphics[scale=0.070]{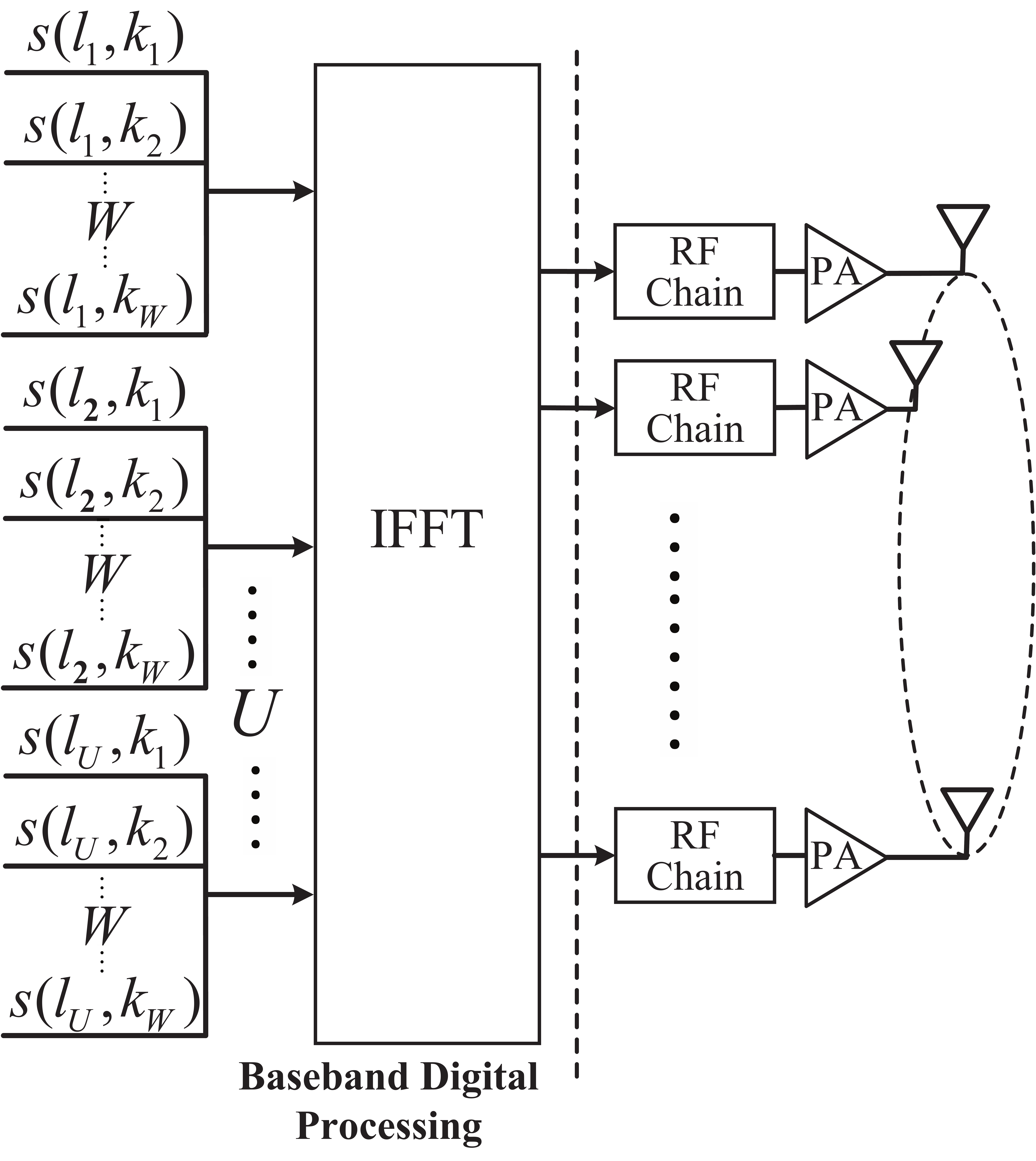}
\caption*{\footnotesize(a)}
\end{minipage}
\hspace{15mm}
\begin{minipage}{5.3cm}%
\includegraphics[scale=0.070]{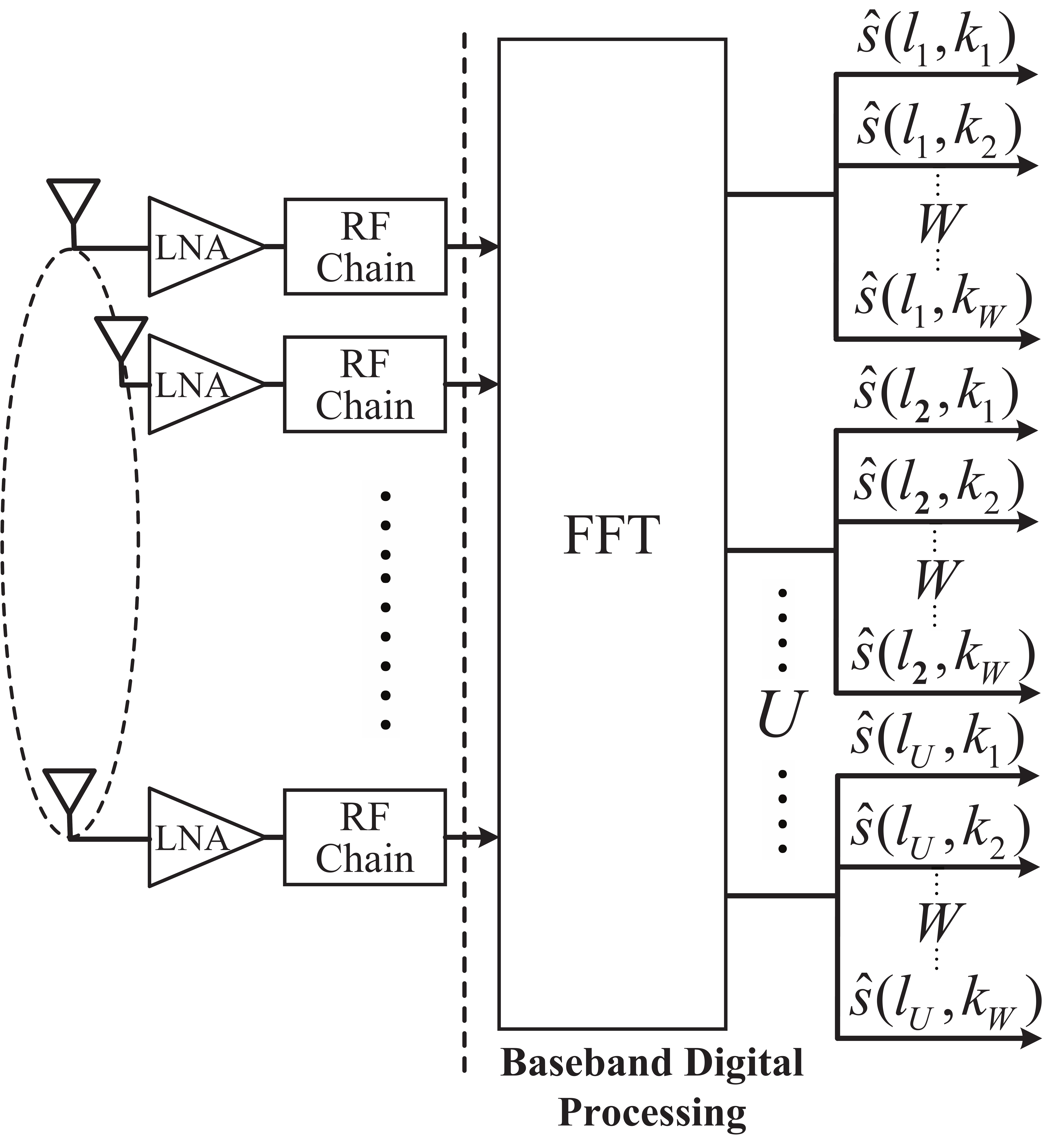}
\caption*{\footnotesize(b)}
\end{minipage}
\caption{The transceiver structure of OAM-OFDM for the downlink MU-OAM transmission. PA: the power amplifier. LNA: low noise
amplifier. (a) The transmitter structure. (b) The receiver structure.}
\label{Fig4}
\setlength{\abovecaptionskip}{0.0cm}   
\setlength{\belowcaptionskip}{-0.2cm}   
\end{figure*}

Denote $\phi_n$ as the azimuth angle of the $n$th $(1\leq n\leq N)$ UCA element at MBS and $\alpha^p_m$ as the azimuth angle of the $m$th ($1\le m \le M$)  UCA element at the $p$th SBS. According to the geometric relationship in Fig.\ref{Fig3}, the transmission distance $d^p_{m,n}$ from the $n$th antenna element of MBS to the $m$th antenna element of the $p$th SBS can be calculated as
\begin{align} \label{dmn}
&d^p_{m,n}=\big[R_t^2+R^2_{r_p}+r_p^2+2r_pR_{r_p}\sin\theta_p\cos(\varphi_p-\alpha^p_m) \nonumber\\
&-2r_pR_t\sin\theta_p\cos(\varphi_p-\phi_n)-2R_tR_{r_p}\cos(\alpha^p_m-\phi_n)]^{1/2},
\end{align}
where $R_t$ and $R_{r_p}$ are respectively the radii of UCAs at MBS and $p$th SBS, $\phi_n=[\frac{2\pi(n-1)}{N}+\phi_0]$, $n=1,2,\cdots,N$, $\alpha^p_m=[\frac{2\pi(m-1)}{M}+\alpha_0]$, $m=1,2,\cdots,M$, $\phi_0$ and $\alpha_0$ are respectively the corresponding initial angles of the first reference antenna elements in both UCAs. For easier analysis, we assume $\phi_0=0$ and $\alpha_0=0$ in this paper. By assuming that all the $P$ SBSs are located in the far-field distance region of MBS, i.e. $r_p\gg R_t$ and $r_p\gg R_{r_p}$, $p=1,2,\cdots,P$, we can approximate $d^p_{m,n}$ as
\begin{align} \label{dmn appx}
&d^p_{m,n}\overset{(a)}{\approx} \sqrt{R_t^2+R^2_{r_p}+r_p^2}+\frac{r_pR_{r_p}\sin\theta_p\cos(\varphi_p-\alpha^p_m)}{\sqrt{R_t^2+R^2_{r_p}+r_p^2}}\nonumber\\
&\quad \quad \quad -\frac{r_pR_t\sin\theta_p\cos(\varphi_p-\phi_n)}{\sqrt{R_t^2+R^2_{r_p}+r_p^2}}-\frac{R_tR_{r_p}\cos(\alpha^p_m-\phi_n)}{\sqrt{R_t^2+R^2_{r_p}+r_p^2}}\nonumber\\
&\overset{(b)}{\approx} r_p+R_{r_p}\sin\theta_p\cos(\varphi_p-\alpha^p_m)-R_t\sin\theta_p\cos(\varphi_p-\phi_n)\nonumber\\
&\quad -\frac{R_tR_{r_p}}{r_p}\cos(\alpha^p_m-\phi_n),
\end{align}
where (a) uses the method of completing a square and the condition $r_p\gg R_t, R_{r_p}$ as same as the simple case $\sqrt{a^2-2b}\approx a-\frac{b}{a}, a\gg b$; (b) is directly obtained from the condition $r_p\gg R_t, R_{r_p}$. Then, substituting \eqref{dmn appx} into \eqref{FreeSpaceChannel} and abbreviating $h(k,d^p_{m,n})$ to $h^p_{m,n}(k)$, we thus have
\begin{align} \label{hmn}
&h_{m,n}^p(k) \overset{(a)}\approx \frac{\beta}{2k r_p}\exp\bigg(-ikr_p-i{k R_{r_p}}\sin\theta_p\cos(\varphi_p-\alpha^p_m) \nonumber\\
&+i{k R_t}\sin\theta_p\cos(\varphi_p-\phi_n)+i\frac{kR_tR_{r_p}}{r_p}\cos(\alpha^p_m-\phi_n) \bigg),
\end{align}
where (a) neglects a few small terms in the denominator and thus only $2k r_p$ is left.

In this way, the channel matrix from MBS to $P$ SBSs can be expressed as $\mathbf{H}(k)=[h_{m,n}^p(k)]_{PM\times N}$. In the proposed MU-OAM system, MBS is required to simultaneously transmit the $P$ group OAM-based signals to $P$ SBSs in the multi-user downlink channel, thus, the $N~(N=PM)$ elements of UCA at MBS should be divided into $P$ groups and each group contains $M$ elements. Therefore, the channel matrix $\mathbf{H}(k)$ can be rewritten as
\begin{align} \label{channel}
&\mathbf{H}(k) =
\begin{small}
\begin{bmatrix}
\mathbf{H}_{1,1}(k) & \mathbf{H}_{1,2}(k) &\cdots &  \mathbf{H}_{1,P}(k)\\
\mathbf{H}_{2,1}(k) & \mathbf{H}_{2,2}(k) &\cdots &  \mathbf{H}_{2,P}(k)\\
\vdots    &   \vdots    &\ddots &  \vdots                \\
\mathbf{H}_{P,1}(k) & \mathbf{H}_{P,2}(k) &\cdots &  \mathbf{H}_{P,P}(k)\\
\end{bmatrix}
\end{small},
\end{align}
where $\mathbf{H}_{p,q}(k)$ is the $M\times M$ submatrix that denote the channel from the $q$th $(1 \leq q \leq P)$ group elements of MBS to the $p$th SBS, which can be written as
\begin{align} \label{subchannel}
&\mathbf{H}_{p,q}(k) = \nonumber \\
&\begin{bmatrix}
h^p_{1,q}(k) & h^p_{1,q+P}(k) &\cdots &  h^p_{1,q+P\times(M-1)}(k)\\
h^p_{2,q}(k) & h^p_{2,q+P}(k) &\cdots &  h^p_{2,q+P\times(M-1)}(k)\\
\vdots    &   \vdots    &\ddots &  \vdots                \\
h^p_{M,q}(k) & h^p_{M,q+P}(k) &\cdots &  h^p_{M,q+P\times(M-1)}(k)\\
\end{bmatrix}.
\end{align}

\vspace{-0.0cm}
\subsection{Signal Model}
\vspace{0.0cm}

For higher data rate transmission, OFDM is often applied and naturally the OAM-OFDM communication system has been proposed in \cite{Chen2018A}. In the downlink MU-OAM wireless backhaul system, we assume $W$ subcarriers and $U$ OAM modes for data transmission, and $\widetilde{W}$ subcarriers and $\widetilde{U}$ OAM modes for training, $1\leq\widetilde{W}\leq W, 1\leq\widetilde{U}\leq U$. The transceiver structure of OAM-OFDM for downlink MU-OAM transmission is illustrated in Fig.\ref{Fig4}.

\vspace{0.2cm}
\subsubsection{Uplink Training Stage}
In the uplink training stage, pilot data is exploited for performing channel estimation. First, MBS broadcast a message to $P$ SBSs. After receiving the broadcast signal, $P$ SBSs simultaneously send to MBS the same OAM-based training symbols known to MBS.

As a UCA can successively generate different mode OAM beams by the baseband (partial) inverse fast Fourier transforms (IFFT), which can be regarded as a precoding in the beam-space, the equivalent baseband training signal model of each SBS can be expressed as $\mathbf{F}_{\widetilde{U}}\mathbf{S'}(k_w)$, where $\mathbf{F}_{\widetilde{U}}$ $=[\mathbf{f}^\mathrm{H}(\ell_1)$,$\mathbf{f}^\mathrm{H}(\ell_2)$,$\cdots$,$\mathbf{f}^\mathrm{H}(\ell_{\widetilde{U}})]$ is a $M\times \widetilde{U}$ (partial) IFFT matrix, $\mathbf{f}(\ell_u)=[1,e^{-i\frac{2\pi \ell_u}{M}},\cdots,$ $e^{-i\frac{2\pi \ell_u(M-1)}{M}}]$,  $\mathbf{S}'(k_w)$ $=\textrm{diag}\{s'(\ell_1,k_w),s'(\ell_2,k_w),\cdots,s'(\ell_{\widetilde{U}},k_w)\}$ is the $\widetilde{U}$-dimensional training symbol matrix transmitted by each SBS, and $s'(\ell_{u},k_w)$ is the $u$th $(1\leq u\leq \widetilde{U})$ OAM mode training symbol transmitted on the $u$th OFDM symbol at the $w$th $(1\leq w\leq \widetilde{W})$ subcarrier. Correspondingly, the received uplink baseband training symbol matrix $\mathbf{Y'}_t(k_w)$ at MBS can be written as
\begin{align} \label{Yp}
\mathbf{Y'}_t(k_w)=\sum_{p=1}^{P}\mathbf{H}^\mathrm{T}_p(k_w)\mathbf{F}_{\widetilde{U}} \mathbf{S'}(k_w)+\mathbf{Z'}_t(k_w),
\end{align}
where $\mathbf{Y'}_t(k_w)$ is the $N\times \widetilde{U}$ matrix, $\mathbf{H}_p(k_w)=[h^p_{m,n}(k_w)]_{M\times N}$ is the channel matrix from MBS to the $p$th SBS at the $w$th subcarrier, and $\mathbf{Z}'_t(k_w)$ is the related additive noise matrix.

Then, the received uplink signals $\mathbf{Y'}_t(k_w)$ on all the $N$ elements of MBS are combined at each OFDM symbol, which can be formulated as
\begin{align} \label{xpie}
\mathbf{x'}_t(k_w)&=\mathbf{1}^\mathrm{T}\mathbf{Y}'_t(k_w)\nonumber \\
&=\mathbf{1}^\mathrm{T}\sum_{p=1}^{P}\left(\mathbf{H}^\mathrm{T}_p(k_w)\mathbf{F}_{\widetilde{U}} \mathbf{S'}(k_w)+\mathbf{Z'}_t(k_w)\right),
\end{align}
where $\mathbf{x}'_t(k_w)$ $=$ $[x'_t(\ell_1,k_w),$ $x'_t(\ell_2,k_w),$ $\cdots,$ $x'_t(\ell_{\widetilde{U}},k_w)]$, $x'_t(\ell_u,k_w)$ is the combined signal corresponding to $s'(\ell_u,k_w)$. Furthermore, how to perform channel estimation at MBS by the received uplink training symbols will be specified in Section III.

\vspace{0.2cm}
\subsubsection{Downlink Data Transmission Stage}
\begin{figure*}[t]
\setcounter{equation}{10}
\begin{align} \label{xtqute}
x'_t(\ell_{u},k_w)&=\sum_{n=1}^{N}\left(\sum_{p=1}^{P}\mathbf{h}^p_n(k_w)\mathbf{f}^\mathrm{H}(\ell_u)s'(\ell_{u},k_w)\right)+z'_t(\ell_{u},k_w) \nonumber\\
&=\frac{\beta}{2k_w}\sum_{p=1}^{P}\sum_{n=1}^{N}\sum_{m=1}^{M} \frac{e^{i\bm{k}_w|\bm{d}^p_{n,m}|}}{|\bm{d}^p_{n,m}|}e^{i\ell_{u}\alpha^p_m}
 s'(\ell_{u},k_w)+z'_t(\ell_{u},k_w)\nonumber\\
&=\frac{\beta}{2k_w}\sum_{p=1}^{P}\sum_{n=1}^{N}\sum_{m=1}^{M} \frac{e^{i\bm{k}_w|\bm{r}_p-\bm{r}'^p_m+\bm{r}'_n|}} {|\bm{r}_p-\bm{r}'^p_m+\bm{r}'_n|}e^{i\ell_{u}\alpha^p_m}
 s'(\ell_{u},k_w)+z'_t(\ell_{u},k_w)\nonumber\\
&\approx\frac{\beta}{2k_w}\sum_{p=1}^{P}\frac{e^{ik_wr_p}}{r_p}\sum_{n=1}^{N} e^{i\bm{k}_w\cdot\bm{r}_n}
\sum_{m=1}^{M} e^{-i(\bm{k}_w\cdot\bm{r}^p_m-\ell_{u}\alpha^p_m)}s'(\ell_{u},k_w) +z'_t(\ell_{u},k_w)\nonumber\\
&\approx\sigma(\ell_u,k_w)
\sum_{p=1}^{P}\frac{e^{ik_wr_p}}{r_p}e^{i\ell_{u}\varphi_p}{J_{\ell_{u}}}(k_wR_{r_p}\sin\theta_p){J_0}(k_wR_t\sin\theta_p)+z'_t(\ell_{u},k_w),
\end{align}
\setcounter{equation}{8}%
\hrulefill
\setlength{\abovecaptionskip}{0cm}   
\setlength{\belowcaptionskip}{0.0cm}   
\end{figure*}
%

As a UCA can also simultaneously generate multi-mode OAM beams with the baseband (partial) IFFT, the equivalent baseband signal model of multi-mode OAM data symbols transmitted by MBS can be expressed $\mathbf{F}\mathbf{s}(k_w)$ , where $\mathbf{F}$ $=$ $\mathbf{I}_P\otimes\mathbf{F}_U$ is the $P$-dimensional block diagonal matrix used to simultaneously generate multi-mode OAM beams, $\mathbf{F}_U=[\mathbf{f}^\mathrm{H}(\ell_1)$, $\mathbf{f}^\mathrm{H}(\ell_2)$, $\cdots$, $\mathbf{f}^\mathrm{H}(\ell_{U})]$ is a $M\times U$ (partial) IFFT matrix and $\otimes$ denotes Kronecker product, $\mathbf{s}^{\mathrm{T}}(k_w)$ $=$ $[\mathbf{s}^\mathrm{T}_1(k_w),$ $\mathbf{s}^\mathrm{T}_2(k_w),$ $\cdots,$ $\mathbf{s}^\mathrm{T}_{P}(k_w)]$ contains all the data symbols transmitted to $P$ SBSs, and $\mathbf{s}^\mathrm{T}_p(k_w)$ $=$ $[s_p(\ell_1,k_w)$,$s_p(\ell_2,k_w)$,$\cdots$,$s_p(\ell_U,k_w)]$ is the data symbol vector transmitted to the $p$th $(1\leq p\leq P)$ SBS on $U$ OAM modes at the $w$th subcarrier. Besides, the interferences in the MU-OAM channel are considered to be eliminated at MBS, thus, the data symbol vector $\mathbf{s}(k_w)$ needs to be preprocessed before performing baseband (partial) IFFT, i.e., $\mathbf{F}\mathbf{P}(k_w)\mathbf{s}(k_w)$, where $\mathbf{P}(k_w)$ is the $PU\times PU$ preprocessing matrix at the $w$th subcarrier.

Correspondingly, the received downlink signal vector at the $p$th SBS can be written as
\begin{align} \label{yp}
\mathbf{y}_{p}(k_w)=\mathbf{H}_{p}(k_w)\mathbf{F}\mathbf{P}(k_w)\mathbf{s}(k_w)+\mathbf{z}_p(k_w),
\end{align}
where $\mathbf{z}_p(k_w)$ is the noise vector.
The UCA-based OAM receiver has the similar baseband digital structure to the transmitter (see Fig.\ref{Fig4} and Fig.3 in \cite{Chen2018A}) but with the opposite operations, i.e., separating different OAM modes and despiralizing each mode. Thus, the detected OAM data symbol vector $\mathbf{x}_{p}(k_w)$ at the $p$th SBS can be expressed as
\begin{align} \label{xp}
\mathbf{x}_{p}(k_w)=\mathbf{F}^\mathrm{H}_U \big(\mathbf{H}_{p}(k_w)\mathbf{F}\mathbf{P}(k_w)\mathbf{s}(k_w)+\mathbf{z}_p(k_w)\big),
\end{align}
where $\mathbf{x}^\mathrm{T}_p(k_w)$ $=$ $[x_p(\ell_1,k_w)$, $x_p(\ell_2,k_w)$, $\cdots,$ $x_p(\ell_U,k_w)]$, $x_p(\ell_u,k_w)$ is the detected data symbol of the $p$th SBS on the $u$th $(1\leq u\leq U)$ OAM mode at the $w$th $(1\leq w\leq W)$ subcarrier. Furthermore, how to design the preprocessing matrix $\mathbf{P}(k_w)$ will be specified in Section V.

\vspace{0.2cm}
\section{Multi-User Distance and AoA Estimation}
\vspace{0.4cm}

In this section, we propose an effective OAM-based multi-user distance and AoA estimation method, which forms the basis of the following MU-OAM preprocessing.

\vspace{0.0cm}
\subsection{Problem Formulation}
\vspace{0.0cm}

We can observe from \eqref{hmn} and \eqref{channel} that if $k_w$ $(1\leq w\leq\widetilde{W}\leq W)$, $R_{r_p}$ $(1\leq p\leq P)$ and $R_t$ are known to MBS, the channel matrix from MBS to $P$ SBSs is only characterized by the position of each SBS. Accordingly, pilot data should be exploited for distances and AoAs estimation of $P$ SBSs during the training stage.

According to \eqref{xpie}, the combined signal $x'_t(\ell_{u},k_w)$ can be derived in \eqref{xtqute}, where $\mathbf{h}^p_n(k_w)$ $=$ $[h^p_{n,1}(k_w),$ $h^p_{n,2}(k_w),$ $\cdots,$ $h^p_{n,M}(k_w)]$, $\bm{k}_w$ is the wave vector, $\bm{d}^p_{n,m}$ is the position vector from the $m$th antenna element of $p$th SBS to the $n$th antenna element of MBS, $\sigma(\ell_u,k_w)=\frac{MN\beta i^{-\ell_u}}{2k_w}s'(\ell_{u},k_w)$, $J_{\ell_u}(\cdot)$ is $\ell_u$th-order Bessel function of the first kind, and $z'_t(\ell_{u},k_w)$ is the noise. In far-field, $|\bm{r}_p-\bm{r}'^p_m+\bm{r}'_n|\approx r_p$ for amplitudes and $|\bm{r}_p-\bm{r}'^p_m+\bm{r}'_n|\approx r_p-\bm{\hat{r}}_p\cdot\bm{r}^p_m+\bm{\hat{r}}_p\cdot\bm{r}_n$ for phases, where $\bm{r}_p$ is the position vector from the UCA center of the $p$th SBS to the UCA center of MBS, $\bm{\hat{r}}_p$ is the unit vector of $\bm{r}_p$, $\bm{r}^p_m=R_{r_p}(\bm{x}_p\cos\alpha^p_m+\bm{y}_p\sin\alpha^p_m)$, $\bm{r}_n=R_t(\bm{x}\cos\phi_n+\bm{y}\sin\phi_n)$, $\bm{x}_p$, $\bm{y}_p$, $\bm{x}$ and $\bm{y}$ are the unit vectors of $\textrm{X}_{p}$-axis, $\textrm{Y}_{p}$-axis, $\textrm{X}$-axis and $\textrm{Y}$-axis respectively.

Overall, the aim of the OAM-based multi-user distance and AoA estimation is to obtain the distance, the azimuthal angle and the elevation angle of each SBS's UCA center from the signals $\{x'_t(\ell_{u},k_w)|{u}=1,2,\dots,{\widetilde{U}}; w=1,2,\cdots,\widetilde{W}\}$.

\vspace{0.0cm}
\subsection{Distance and AoA Estimation}
\vspace{0.0cm}

Since the training signals of multiple SBSs have different amplitudes arriving at MBS, it is difficult to directly extract the exponentials containing $r_p$ and $\varphi_p$ from \eqref{xtqute} as in \cite{Chen2020Multi,Long2021AoA}. Hence, the 2-D ESPRIT-based distance and AoA estimation method proposed for SU-OAM scenario\cite{Chen2020Multi,Long2021AoA} is not suitable for MU-OAM scenario. Inspired by the transform domain techniques in OAM radar imaging \cite{Kang2015Orbital}, we propose to first estimate $\{(r_p,\varphi_p)|p=1,2,\cdots,P\}$, and then estimate $\{\theta_p|p=1,2,\cdots,P\}$ based on \eqref{xtqute} with the FFT/two-dimensional (2-D) FFT method.

\vspace{0.2cm}
\subsubsection{Estimation of $r_p$ and $\varphi_p$}
The combined signal $x'_t(\ell_{u},k_w)$ in \eqref{xtqute} can be simplified as
\setcounter{equation}{11}
\begin{align}
&\tilde{x}'_t(\ell_{u},k_w)=\frac{x'_t(\ell_{u},k_w)}{|\sigma(\ell_{u},k_w)|} \frac{s'(\ell_{u},k_w)^*}{|s'(\ell_{u},k_w)|}i^{\ell_u}\nonumber\\
&=\sum_{p=1}^{P}\frac{e^{ik_wr_p}}{r_p}e^{i\ell_{u}\varphi_p}{J_{\ell_{u}}}(k_wR_{r_p}\sin\theta_p){J_0}(k_wR_t\sin\theta_p)\nonumber\\
&\quad +\tilde{z}'_t(\ell_{u},k_w),
\label{xtuw}
\end{align}
where $\tilde{z}'_t(\ell_{u},k_w)$ is the noise. Then, all the signals received on the ${\widetilde{U}}$ OAM modes at the  $\widetilde{W}$ subcarriers can be collected in the matrix
\begin{equation}\label{matrixX}
\mathbf{\widetilde{X}}=
\begin{bmatrix}
\tilde{x}'_t(\ell_1,k_1)               & \tilde{x}'_t(\ell_1,k_2)                  & \cdots & \tilde{x}'_t(\ell_1,k_{\widetilde{W}}) \\
\tilde{x}'_t(\ell_2,k_1)               & \tilde{x}'_t(\ell_2,k_2)                  & \cdots & \tilde{x}'_t(\ell_2,k_{\widetilde{W}}) \\
    \vdots                             &   \vdots                                  & \ddots &   \vdots   \\
\tilde{x}'_t(\ell_{\widetilde{U}},k_1) & \tilde{x}'_t(\ell_{\widetilde{U}},k_2)    & \cdots &\tilde{x}'_t(\ell_{\widetilde{U}},k_{\widetilde{W}}) \\
\end{bmatrix}.
\end{equation}
From \eqref{xtuw}, we can find that the OAM mode $\ell_u$ and the azimuth angle $\varphi_p$ are related through the exponential term $e^{i\ell_{u}\varphi_p}$. Similarly, the subcarrier $k_w$ and the distance $r_p$ are related through the exponential term $e^{ik_w r_p}$. According to the Fourier transform, $\ell_u$ and $\varphi_p$ satisfy the dual relationship, so do $k_w$ and $r_p$, which means that $\{(r_p,\varphi_p)|p=1,2,\cdots,P\}$ can be estimated directly by 2-D FFT for \eqref{matrixX} in frequency domain and OAM mode domain \cite{Kang2015Orbital}.

\vspace{-0.0cm}
\subsubsection{Estimation of $\theta_p$}
\vspace{0.0cm}

According to \eqref{Yp} and \eqref{xtqute}, the training signal received by $n$th $(1\leq n \leq N)$ element of MBS on the zero OAM mode at $w$th subcarrier can be written as
\begin{align} \label{ytn}
y^n_t(k_w)\approx&\sigma'(k_w)\sum_{p=1}^{P}\frac{e^{ik_wr_p}}{r_p}e^{ik_wR_t\sin\theta_p\cos(\varphi_p-\phi_n)}\nonumber\\
&\times{J_0}(k_wR_{r_p}\sin\theta_p)+z^n_t(k_w),
\end{align}
where $\sigma'(k_w)=\frac{M\beta}{2k_w}s'(\ell_0,k_w)$, $z^n_t(k_w)$ is the related noise. Similarity, the signal $y^n_t(k_w)$ can be simplified as
\begin{align}
\tilde{y}^n_t(k_w)&=\frac{y^n_t(k_w)}{|\sigma'(k_w)|} \frac{s'(\ell_0,k_w)^*}{|s'(\ell_0,k_w)|}\nonumber\\
&=\sum_{p=1}^{P}\frac{1}{r_p}{J_0}(k_wR_{r_p}\sin\theta_p)e^{ik_w\xi^n_p} +\tilde{z}^n_t(k_w),
\label{ytnkw}
\end{align}
where $\xi_p^n=r_p+R_t\sin\theta_p\cos(\varphi_p-\phi_n)$, and $\tilde{z}^n_t(k_w)$ is the noise. Then, the $\widetilde{W}$-dimensional vector $\tilde{\mathbf{y}}^n_t(k_w)$ can be constructed as
\begin{equation}\label{ywtnkw}
\tilde{\mathbf{y}}^n_t(k_w)=[\tilde{y}^n_t(k_1), \tilde{y}^n_t(k_2),\cdots, \tilde{y}^n_t(k_{\widetilde{W}})]^\mathrm{T}.
\end{equation}
We can find from \eqref{ytnkw} that the wave number $k_w$ and the parameter $\xi_p^n$ satisfy the dual relationship. Similarity, FFT can be applied to \eqref{ywtnkw} to estimate $\{\xi_p^n|p=1,2,\cdots, P\}$. After that, according to
\begin{align}\label{est_pro}
\hat{\theta}^n_{q}=\arcsin\left(\frac{\hat{\xi}^n_p-\hat{r}_q}{R_t\cos(\hat{\varphi}_q-\phi_n)}\right), q=1,2,\cdots,P,
\end{align}
the $P$ possible estimates of $\theta_p$ can be obtained, where $\hat{r}_p$, $\hat{\varphi}_p$ and $\hat{\xi}^n_p$ are the estimates of $r_p$, $\varphi_p$ and $\xi^n_p$ respectively.

Thereafter, we need to match $\{(\hat{r}_p,\hat{\varphi}_p)|$ $p$ $=$ $1,$ $2,$ $\cdots,$ $P\}$ to $\{\hat{\theta}^n_{q}|$ $q$ $=$ $1,$ $2,$ $\cdots,$ $P^2\}$. First, we substitute $\{(\hat{r}_p,\hat{\varphi}_p)|$ $p=$ $1,$ $2,$ $\cdots,$ $P\}$ and $\{\hat{\theta}^n_{q}|q$ $=$ $1,$ $2,$ $\cdots,$ $P^2\}$ into $e^{ik_w \hat{r}_p}e^{ik_wR_t\sin\hat{\theta}^n_{q}\cos(\hat{\varphi}_p-\phi_n)}$, and then find a set of solutions $\{(\hat{r}_p,\hat{\theta}^n_p,\hat{\varphi}_p)|p=1,2,\cdots,P\}$ that satisfy \eqref{ytn}, where $\hat{\theta}^n_p$ is adopted as the estimate of $\theta_p$ corresponding to $\tilde{\mathbf{y}}^n_t(k_w)$ $(1\leq n \leq N)$. Hence, there are $N$ estimates of $\theta_p$ in total, which can be expressed as
\begin{align}
\hat{\theta}_p^n=&\theta_p+\varepsilon_p^n,  n=1,2,\cdots,N,\nonumber
\end{align}
where $\{\varepsilon_p^n|n$$=$$1,2,\cdots,N\}$ represent the estimation errors. Assume $\{\varepsilon_p^n|n=1,2,\cdots,N\}$ have the same average variance $\textrm{Var}(\varepsilon_{\theta})$. Thus,
\begin{align}
\textrm{Var}\left(\frac{1}{N}\sum_{n=1}^{N}\hat{\theta}^n_p\right)= \frac{\textrm{Var}\left(\varepsilon_{\theta}\right)}{N}.
\end{align}
Finally, $\hat{\theta}_p=\frac{1}{N}\sum_{n=1}^{N}\hat{\theta}_p^n$ is adopted as the estimate of $\theta_p$, and the OAM-based multi-user distance and AoA estimation is achieved. Since the signals in \eqref{matrixX} and \eqref{ywtnkw} are coherent but the corresponding noises are incoherent, only the signals are enhanced when (2-D) FFT is conducted. As we can see in the simulation results in Section VI, the noises have little influence on the OAM-based multi-user distance and AoA estimation when using (2-D) FFT method.

\vspace{0.0cm}
\section{Preprocessing of Downlink MU-OAM System}
\vspace{0.0cm}

In this section, we first apply a MU-OAM preprocessing scheme for the proposed system, which can effectively eliminate the co-mode and inter-mode interferences in the downlink MU-OAM channel. After that, we analyze SE and EE performances of the downlink MU-OAM wireless backhaul system.

\vspace{-0.0cm}
\subsection{Problem Formulation}
\vspace{0.0cm}

From \eqref{xp}, the effective OAM channel matrix from MBS to $P$ SBSs at the $w$th subcarrier can be expressed as
\begin{align} \label{effectivechannel}
\mathbf{H}_{\rm OAM}(k_w)=\mathbf{F}^\mathrm{H}\mathbf{H}(k_w)\mathbf{F}=[\mathbf{H}^{p,q}_{\rm OAM}(k_w)]_{P\times P},
\end{align}
where $\mathbf{H}^{p,q}_{\rm OAM}(k_w)= \mathbf{F}^\mathrm{H}_U\mathbf{H}_{p,q}(k_w)\mathbf{F}_U$ is the $U\times U$ effective OAM channel matrix corresponding to $\mathbf{H}_{p,q}(k_w)$, and the expression of the elements in $\mathbf{H}^{p,q}_{\rm OAM}(k_w)$ can be written as
\begin{align*}
&h^{p,q}_{\textrm{OAM}}(u,v)=\sum\limits_{m = 1}^M \sum\limits_{\bar{n} = 1}^M {h^p_{m,q+(\bar{n}-1)\times P}}\exp \left(-i{\ell_u}\alpha^p_m+i{\ell_v}\phi'_{\bar{n}} \right) \nonumber\\
\end{align*}
\begin{align} \label{heffDev}
&=\delta_{p}(k_w) \sum\limits_{m = 1}^M \sum\limits_{\bar{n} = 1}^M \exp \bigg( -i{\ell_u}\alpha^p_m+i{\ell_v}\phi'_{\bar{n}}\nonumber\\
&\quad +i\frac{k_wR_tR_{r_p}}{r_p}\cos\left(\alpha^p_m-\phi_{\bar{n}}\right)
+i{k_w R_t}\sin\theta_p\cos\left(\varphi_p-\phi_{\bar{n}}\right)\nonumber \\
&\quad -i{k_w R_{r_p}}\sin\theta_p\cos(\varphi_p-\alpha^p_m)
\bigg),
\end{align}
where $\delta_{p}(k_w)$ $=$ $\frac{\beta}{2k_w r_p}\exp\left(-ik_wr_p\right)$, $\phi'_{\bar{n}}$ $=$ $\frac{2\pi(\bar{n}-1)}{M}$, and $\phi_{\bar{n}}$ $=$ $\frac{2\pi[(\bar{n}-1)\times P+q-1]}{N}$. Given that $k_w$, $R_{r_p}$, $R_t$, $M$ and $\ell_u$ are known to MBS,
the effective channel coefficient $h^{p,q}_{\textrm{OAM}}(u,v)$ is only the function of $r_p$, $\varphi_p$ and $\theta_p$. Hence, MBS can obtain the full information of $\mathbf{H}_{\rm OAM}(k_w)$ after OAM-based multi-user distance and AoA estimation. Naturally, the preprocessing process should be performed at MBS to eliminate interferences in the downlink MU-OAM channel.

Denote the effective OAM channel matrix from MBS to the $p$th SBS as $\mathbf{H}^{p}_{\rm OAM}(k_w)$ $=$ $\mathbf{F}^\mathrm{H}_U \mathbf{H}_{p}(k_w)\mathbf{F}$ and the associated preprocessing matrix as $\mathbf{P}_p(k_w)$. Since the role of $\mathbf{P}_p(k_w)$ is to eliminate the co-mode interference from other users and the inter-mode interference in $\mathbf{H}^{p}_{\rm OAM}(k_w)$, $\mathbf{H}^p_{\rm OAM}(k_w)\mathbf{P}_q(k_w)=0$ $(p\neq q)$ and $\mathbf{H}^p_{\rm OAM}(k_w)\mathbf{P}_p(k_w)$  should be the diagonal matrix with $\mathbf{H}^{p}_{\rm OAM}(k_w)$ being the $U\times PU$ matrix and $\mathbf{P}_p(k_w)$ being the $PU\times U$ matrix.

\vspace{0.0cm}
\subsection{Preprocessing for Interference Elimination in Downlink MU-OAM System}
\vspace{0.0cm}

\subsubsection{Co-mode Interference Elimination}
We first construct the co-mode interference elimination matrix as
\begin{align} \label{comodematrix}
\mathbf{E}(k_w)=[\mathbf{E}_1(k_w),\cdots,\mathbf{E}_p(k_w),\cdots,\mathbf{E}_P(k_w)],
\end{align}
where $\mathbf{E}_p(k_w)$ is the $PU\times U$ submatrix for the $p$th SBS at the $w$th subcarrier. To eliminate the co-mode interference from other users, the constraint $\mathbf{H}^p_{\rm OAM}(k_w)\mathbf{E}_q(k_w)=0$ for $p\neq q$ should be satisfied. That is, $\mathbf{E}_p(k_w)$ should lie within the null space of $\widehat{\mathbf{H}}^p_{\rm OAM}(k_w)$, where $\widehat{\mathbf{H}}^p_{\rm OAM}(k_w)$ is defined as
%
\begin{align} \label{systemchannelnotj}
\widehat{\mathbf{H}}^p_{\rm OAM}(k_w)=
\begin{small}
\begin{bmatrix}
\mathbf{H}^1_{\rm OAM}(k_w)\\
\vdots \\
\mathbf{H}^{p-1}_{\rm OAM}(k_w)\\
\mathbf{H}^{p+1}_{\rm OAM}(k_w)\\
\vdots \\
\mathbf{H}^P_{\rm OAM}(k_w)
\end{bmatrix}.
\end{small}
\end{align}
Fortunately, the null space of $\widehat{\mathbf{H}}^p_{\rm OAM}(k_w)$ can be computed by the SVD of $\widehat{\mathbf{H}}^p_{\rm OAM}(k_w)$, which can be written as
\begin{align} \label{svdHpOAM}
\widehat{\mathbf{H}}^p_{\rm OAM}(k_w)&=\widehat{\mathbf{U}}_{p,w}\widehat{\mathbf{\Sigma}}_{p,w}\widehat{\mathbf{V}}^\mathrm{H}_{p,w}\nonumber\\
&=\widehat{\mathbf{U}}_{p,w}\widehat{\mathbf{\Sigma}}_{p,w}
\begin{bmatrix}
\widehat{\mathbf{V}}_{p,w}^{(1)}, &\widehat{\mathbf{V}}_{p,w}^{(2)}
\end{bmatrix},
\end{align}
$\widehat{\mathbf{U}}_{p,w}$ is the $(P-1)U\times (P-1)U$ left singular matrix, $\widehat{\mathbf{\Sigma}}_{p,w}$ is the $(P-1)U\times PU$ matrix including all singular values of $\widehat{\mathbf{H}}^p_{\rm OAM}(k_w)$, and $\widehat{\mathbf{V}}^\mathrm{H}_{p,w}$ is the $PU\times PU$ right singular matrix. Denote the rank of $\widehat{\mathbf{H}}^p_{\rm OAM}(k_w)$ as $\hat{L}_p$, then $\widehat{\mathbf{V}}^\mathrm{H}_{p,w}$ can be divided into two parts, where $\widehat{\mathbf{V}}_{p,w}^{(1)}$ holds the first $\hat{L}_p$ right singular vectors, and $\widehat{\mathbf{V}}_{p,w}^{(2)}$ holds the last $(PU-\hat{L}_p)$ right singular vectors. Therefore, $\widehat{\mathbf{V}}_{p,w}^{(2)}$ forms an orthogonal basis for the null space of $\widehat{\mathbf{H}}^p_{\rm OAM}(k_w)$, and its columns are, thus, candidates for $\mathbf{E}_p(k_w)$. As $\hat{L}_p\leq (P-1)U$, $\widehat{\mathbf{V}}_{p,w}^{(2)}$ contains at least $U$ right singular vectors. Thereafter, we choose the last $U$ columns of $\widehat{\mathbf{V}}_{p,w}^{(2)}$ to form $\mathbf{E}_p(k_w)$.

After eliminating the co-mode interference by $\mathbf{E}(k_w)$, the effective OAM channel matrix of downlink MU-OAM wireless backhaul system at the $w$th subcarrier becomes
\begin{align}\label{effectivechannel2}
&\mathbf{H}'_{\rm OAM}(k_w)=\mathbf{H}_{\rm OAM}(k_w)\mathbf{E}(k_w)\nonumber\\
&=\begin{bmatrix}
\begin{small}
\begin{matrix}
\mathbf{H}^1_{\rm OAM}(k_w)\mathbf{E}_1(k_w) & &\mathbf{0}\\
&\ddots &\\
\mathbf{0} & &\mathbf{H}^P_{\rm OAM}(k_w)\mathbf{E}_P(k_w)
\end{matrix}
\end{small}
\end{bmatrix}.
\end{align}
where $\mathbf{H}^p_{\rm OAM}(k_w)\mathbf{E}_p(k_w)$ is the $U\times U$ effective OAM channel matrix from MBS to the $p$th SBS. Comparing \eqref{effectivechannel} and \eqref{effectivechannel2}, we can see that the MU-OAM channel is decoupled to $P$ parallel single-user OAM channels by $\mathbf{E}(k_w)$.

\vspace{0.2cm}
\subsubsection{Inter-mode Interference Elimination}
The angles $\{\varphi_p|p$ $=$ $1,2,\cdots,P\}$ and $\{\theta_p|p$ $=$ $1,2,\cdots,P\}$ between the transmit beam and each receive beam causes inter-mode interference, which results in $\mathbf{H}^p_{\rm OAM}(k_w)\mathbf{E}_p(k_w)$ not being a diagonal matrix \cite{Chen2020Multi}. Thus, the purpose of the inter-mode interference elimination is to remove the off-diagonal elements in $\mathbf{H}^p_{\rm OAM}(k_w)\mathbf{E}_p(k_w)$ so that $\mathbf{H}^p_{\rm OAM}(k_w)\mathbf{E}_p(k_w)$ can be diagonalized.

We construct the inter-mode interference elimination matrix as
\begin{align} \label{G}
&\mathbf{G}(k_w)=\textrm{diag}\{\mathbf{G}_1(k_w),\cdots,\mathbf{G}_p(k_w),\cdots,\mathbf{G}_P(k_w)\},
\end{align}
where $\mathbf{G}_p(k_w)$ is the $U\times U$ submatrix for the $p$th SBS at the $w$th subcarrier. The role of $\mathbf{G}_p(k_w)$ can be viewed as diagonalizing the product $\mathbf{H}^p_{\rm OAM}(k_w)\mathbf{E}_p(k_w)\mathbf{G}_p(k_w)$. Therefore, $\mathbf{G}_p(k_w)$ can be easily formed by the inverse of the product $\mathbf{H}^p_{\rm OAM}(k_w)\mathbf{E}_p(k_w)$, i.e., $\mathbf{G}_p(k_w)=\big(\mathbf{H}^p_{\rm OAM}(k_w)\mathbf{E}_p(k_w)\big)^{-1}$.

\begin{figure*}[t]
\setlength{\abovecaptionskip}{-0cm}   
\setlength{\belowcaptionskip}{-0.3cm}   
\begin{center}
\includegraphics[scale=0.24]{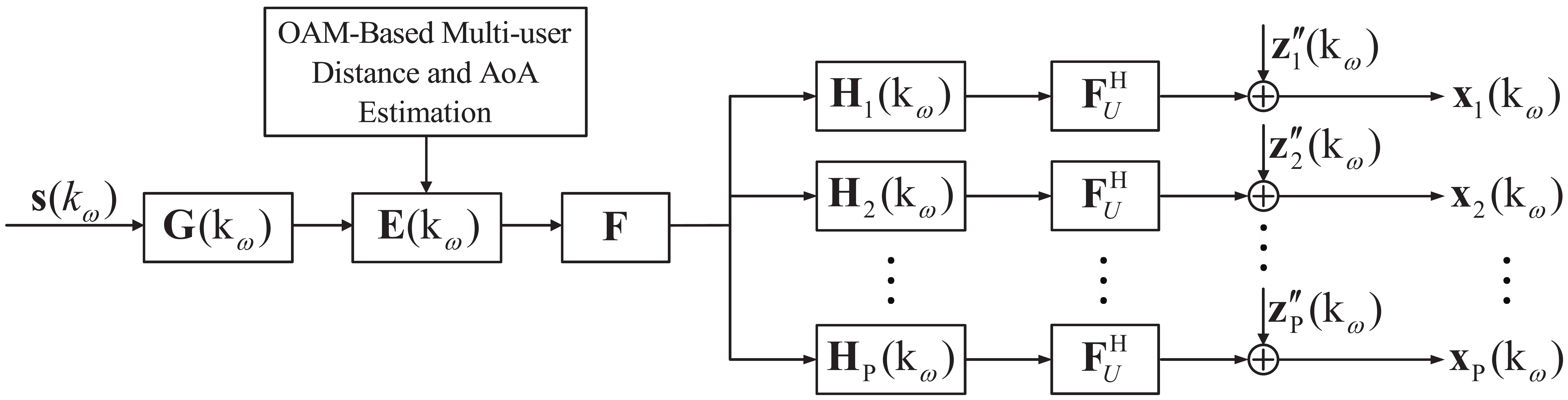}
\end{center}
\caption{The block diagram of a downlink MU-OAM wireless backhaul system with preprocessing assisted by the OAM-based multi-user distance and AoA estimation at MBS.}
\label{Fig5}
\end{figure*}

Through above analysis, the preprocessing matrix $\mathbf{P}_p(k_w)$ can be designed as the cascade of $\mathbf{E}_p(k_w)$ and $\mathbf{G}_p(k_w)$, i.e. $\mathbf{P}_p(k_w)=\mathbf{E}_p(k_w)\mathbf{G}_p(k_w)$. Correspondingly, the preprocessing matrix of the downlink MU-OAM system $\mathbf{P}(k_w)$ can be expressed as
\begin{align} \label{InterferenceEliminationMatrix2}
\mathbf{P}(k_w)&=\mathbf{E}(k_w)\mathbf{G}(k_w)\nonumber\\
&=[\mathbf{E}_1(k_w)\mathbf{G}_1(k_w),\cdots,\mathbf{E}_P(k_w)\mathbf{G}_P(k_w)].
\end{align}
As the effective channel of the downlink MU-OAM system is completely characterized by the position of each SBS, the preprocessing matrix $\mathbf{P}(k_w)$ can be designed only according to the position information of $P$ SBSs instead of all the channel state information (CSI). Therefore, the channel parameters needed by the proposed MU-OAM preprocessing scheme are significantly less than that in the traditional MIMO precoding. If $\{(\hat{r}_p,\hat{\theta}_p,\hat{\varphi}_p)|p$ $=$ $1,2,\cdots,P\}$ obtained by OAM-based multi-user distance and AoA estimation are accurate enough, the effective OAM channel matrix after preprocessing $\mathbf{H}_{\rm OAM}(k_w)\mathbf{P}(k_w)$ approaches diagonal. Fig.\ref{Fig5} shows a block diagram of a UCA-based downlink MU-OAM wireless backhaul system with the preprocessing assisted by the OAM-based multi-user distance and AoA estimation at MBS.

\vspace{0.0cm}
\subsection{Spectrum Efficiency of Downlink MU-OAM System}
\vspace{-0.0cm}

Following the signal model in \eqref{xp},  the detected OAM data symbol vector at the $p$th SBS can be written as
\begin{align} \label{xpkw1'}
\mathbf{x}_{p}(k_w)=&\mathbf{F}^\mathrm{H}_U \big(\mathbf{H}_{p}(k_w)\mathbf{F}\mathbf{P}(k_w)\mathbf{s}(k_w)+\mathbf{z}_p(k_w)\big)\nonumber\\
=&\mathbf{F}^\mathrm{H}_U \big(\mathbf{H}_{p}(k_w)\mathbf{F}\sum_{q=1}^{P}\mathbf{P}_q(k_w)\mathbf{s}_q(k_w)+\mathbf{z}_p(k_w)\big)\nonumber\\
=&\mathbf{H}^p_{\rm OAM}(k_w)\mathbf{P}_p(k_w)\mathbf{s}_p(k_w)\nonumber\\
&+\mathbf{H}^p_{\rm OAM}(k_w)\widehat{\mathbf{P}}_p(k_w)\hat{\mathbf{s}}_p(k_w)+\mathbf{z}''_p(k_w) \nonumber\\
=&\mathbf{s}_p(k_w)+\bm{\mathcal{I}}^p_{inter}(k_w)+\bm{\mathcal{I}}^p_{co}(k_w)+\mathbf{z}''_p(k_w).
\end{align}
where $\widehat{\mathbf{P}}_p(k_w)$ and $\hat{\mathbf{s}}_p(k_w)$ are respectively defined as
\begin{align*}
\widehat{\mathbf{P}}_p(k_w)&\!=\![\mathbf{P}_1(k_w),  \cdots, \mathbf{P}_{p-1}(k_w),\mathbf{P}_{p+1}(k_w), \cdots, \mathbf{P}_P(k_w)],\nonumber\\
\hat{\mathbf{s}}^\mathrm{T}_p(k_w)&\!=\![\mathbf{s}^\mathrm{T}_1(k_w), \cdots, \mathbf{s}^\mathrm{T}_{p-1}(k_w),\mathbf{s}^\mathrm{T}_{p+1}(k_w), \cdots, \mathbf{s}^\mathrm{T}_{P}(k_w)],
\end{align*}
$\bm{\mathcal{I}}^p_{inter}(k_w)$ and $\bm{\mathcal{I}}^p_{co}(k_w)$ are respectively the $U$-dimensional inter-mode interference vector and co-mode interference vector induced by the error in preprocessing matrix, and $\mathbf{z}''_p(k_w)$ is the noise vector.

Since both the co-mode interference and inter-mode interference are eliminated at MBS by the MU-OAM preprocessing, $P$ SBSs can directly obtain detected OAM data symbol vector after separating and despiralizing OAM beams, which can greatly reduce the complexity of each SBS in downlink MU-OAM system. However, considering the inevitable estimation errors of $\{(\hat{r}_p,\hat{\theta}_p,\hat{\varphi}_p)|p=1,2,\cdots,P\}$ in practice, we have to evaluate the effect of these errors on the system SE performance. $\mathbf{R}^p_{\rm inter}(k_w)$, $\mathbf{R}^p_{\rm co}(k_w)$ and $\mathbf{R}^p_z(k_w)$ are defined as the $U\times U$ covariance matrices of $\bm{\mathcal{I}}^p_{inter}(k_w)$, $\bm{\mathcal{I}}^p_{co}(k_w)$ and $\mathbf{z}''_p(k_w)$, i.e.,
\begin{align*}
&\mathbf{R}^p_{\rm inter}(k_w)=\left(\mathbf{H}^p_{\rm OAM}(k_w)\mathbf{P}_p(k_w)-\mathbf{I}_U\right)\mathbb{E}\left(\mathbf{s}_p(k_w)\mathbf{s}_p^\mathrm{H}(k_w)\right)\nonumber\\
&\qquad\qquad\qquad\!\!\left(\mathbf{H}^p_{\rm OAM}(k_w)\mathbf{P}_p(k_w)-\mathbf{I}_U\right)^\mathrm{H}, \nonumber\\
&\mathbf{R}^p_{\rm co}(k_w)=\sum_{q=1,q\neq p}^{P}\!\!\big(\mathbf{H}^p_{\rm OAM}(k_w)\mathbf{P}_q(k_w)\big)
                      \mathbb{E}\left(\mathbf{s}_q(k_w)\mathbf{s}_q^\mathrm{H}(k_w)\right) \nonumber\\
&\qquad\qquad\qquad\qquad\!\!\big(\mathbf{H}^p_{\rm OAM}(k_w)\mathbf{P}_q(k_w)\big)^\mathrm{H}, \nonumber\\
&\mathbf{R}^p_z(k_w)=\mathbb{E}\left(\mathbf{z}''_p(k_w) (\mathbf{z}''_p(k_w))^\mathrm{H}\right),
\end{align*}
where $\mathbf{I}_U$ is the $U$-dimensional unit matrix. Then, the SINR of the $p$th SBS on the $u$th mode OAM at the $w$th subcarrier can be formulated as
\begin{equation*} \label{SINR}
\textrm{SINR}_{p}(\ell_u,k_w)=\frac{\mathbb{E} \left(\left|s_{p}(\ell_u,k_p)\right|^2\right)}{\left[\mathbf{R}^p_{\rm inter}(k_w)\right]_u +
\left[\mathbf{R}^p_{\rm co}(k_w)\right]_u+\left[\mathbf{R}^p_z(k_w)\right]_u},
\end{equation*}
where $[\cdot]_{u}$ denotes the $u$th diagonal element. Therefore, the SE of the UCA-based downlink MU-OAM wireless backhaul system can be written as
\begin{align} \label{CMU}
C=\bigg(1-\frac{T_t\widetilde{W}}{T_cW}\bigg) \frac{1}{W}\sum_{p=1}^P\sum_{w=1}^W\sum_{u=1}^U \log_2\left(1 +\textrm{SINR}_p(\ell_u,k_w)\right),
\end{align}
where $T_c$ is the length of the coherence time over which the channel can be assumed constant, and $T_t$ is the time spent in the training stage.

\vspace{0.0cm}
\subsection{Energy Efficiency of Downlink MU-OAM System}
\vspace{0.0cm}

The EE, defined as the ratio of the system capacity to the total system power consumption, has been recognized as an important performance metric of green communication systems. Due to the high SE of MU-OAM relying on the high transmit power, it is therefore necessary to investigate the EE performance of the MU-OAM wireless backhaul system.

For the proposed MU-OAM system, the consumed total power can be expressed as
\begin{align} \label{Ptotal}
\mathcal{P}_{\textrm{total}} =\frac{\mathcal{P}_{\textrm{T}}}{\rho}+\mathcal{P}_{\textrm{c}},
\end{align}
where $\mathcal{P}_{\textrm{T}}$ is the total transmit power, $\rho$ is the efficiency of the power amplifier (PA), and $\mathcal{P}_{\textrm{c}}$ is the total circuit power consumption which is modeled as a linear function of the number of transmission antennas and RF chains. For easier analysis, we consider equal power allocation at MBS and denote the power of the transmit UCA at each subcarrier as $\mathcal{P}_{\textrm{t}}$, thus $\mathcal{P}_{\textrm{T}}=W\mathcal{P}_{\textrm{t}}$. Besides, for proposed MU-OAM system with the baseband digital structure, the total circuit power consumption includes the powers of the baseband, the RF chains and the low noise amplifiers (LNAs), thus $\mathcal{P}_{\textrm{c}}$ can be written as
\begin{align} \label{Pcdbf}
\mathcal{P}_{\textrm{c}} &= (1+P)\mathcal{P}_{\textrm{BB}}+(N+MP)\mathcal{P}_{\textrm{RF}}+MP\mathcal{P}_{\textrm{LNA}}\nonumber\\
&=(1+P)\mathcal{P}_{\textrm{BB}}+2MP\mathcal{P}_{\textrm{RF}}+MP\mathcal{P}_{\textrm{LNA}},
\end{align}
where $\mathcal{P}_{\textrm{BB}}$, $\mathcal{P}_{\textrm{RF}}$ and $\mathcal{P}_{\textrm{LNA}}$ are the powers consumed by a baseband processor, an RF chain of the transmitter/receiver and a LNA respectively, $\mathcal{P}_{\textrm{RF}}$ consists of the powers consumed by a mixer, a local oscillator, a filter and a DAC/ADC \cite{Lin2016Energy}. Then, the EE of the UCA-based MU-OAM wireless backhaul system can be expressed as
\begin{align}\label{EE}
\eta_{\textrm{EE}} = \frac{BC}{\mathcal{P}_{\textrm{total}}} = \frac{BC}{W\mathcal{P}_{\textrm{t}}/\rho + \mathcal{P}_{\textrm{c}}},
\end{align}
where $B$ is the system available bandwidth.

\vspace{0.0cm}
\section{UCCA-Based downlink MU-OAM-MIMO Wireless Backhaul System}
\vspace{0.0cm}

In this section, we first extend the proposed MU-OAM preprocessing method to the UCCA-based downlink MU-OAM-MIMO wireless backhaul system, which can further increase the JSDCM gain of the MU-OAM communications. After that, we analyze the performance and the computational complexity of the proposed UCCA-based MU-OAM-MIMO wireless backhaul scheme.

\vspace{0.0cm}
\subsection{Downlink MU-OAM-MIMO Wireless Backhaul Systems With UCCAs}
\vspace{0.0cm}
%

To further increase the coaxial multiplexing gain of the UCA-based OAM communication systems, multiple concentric UCAs are exploited in the transmitter and the receiver to achieve 200 Gbps data rate \cite{Yagi2021Gbit}. Inspired by this, we extend the proposed methods to the downlink MU-OAM system combined with MIMO. The transceiver structure of the MU-OAM-MIMO system is similar to that of the MU-OAM system shown in Fig.\ref{Fig4} (i.e, replace UCAs with UCCAs).

For the proposed UCCA-based downlink MU-OAM-MIMO system, we assume that the UCCAs at MBS and $P$ SBSs both consist of $\mathfrak{N}$ concentric UCAs with radii $\{R_{t_\mathfrak{n}}|\mathfrak{n}=1,2,\cdots,\mathfrak{N}\}$ and $\{R_{r_{p,\mathfrak{m}}}|p=1,2,\cdots,P;\mathfrak{m}=1,2,\cdots,\mathfrak{N}\}$. Then, the channel matrix of the UCCA-based downlink MU-OAM-MIMO wireless backhaul system can be written as
\begin{align}\label{HMIMO}
\mathbf{\bar{H}}(k_w)=\left[\mathbf{\bar{H}}^\mathrm{T}_1(k_w),\cdots,\mathbf{\bar{H}}^\mathrm{T}_p(k_w),\cdots,\mathbf{\bar{H}}^\mathrm{T}_P(k_w)\right]^\mathrm{T},
\end{align}
where $\mathbf{\bar{H}}_p(k_w)$ $=$ $\left[\mathbf{H}^p_{\mathfrak{m},\mathfrak{n}}(k_w)\right]_{\mathfrak{N}\times\mathfrak{N}}$, and $\mathbf{H}^p_{\mathfrak{m},\mathfrak{n}}(k_w)$ denotes the $M\times N$ channel matrix from the $\mathfrak{n}$th UCA in UCCA of MBS to the $\mathfrak{m}$th UCA in UCCA of the $p$th SBS, $\mathfrak{m},\mathfrak{n}=1,2,\cdots,\mathfrak{N}$, $p=1,2,\cdots,P$. Similar to UCA-based downlink MU-OAM systems, each UCA in UCCA of MBS in the downlink MU-OAM-MIMO system is also divided into $P$ groups to simultaneously transmit $P$ OAM-based signals to $P$ SBSs, and then $\mathbf{H}^p_{\mathfrak{m},\mathfrak{n}}(k_w)$ can be expressed as
\begin{align}\label{HOAMMIMO}
\mathbf{H}^p_{\mathfrak{m},\mathfrak{n}}(k_w)\!=\![\mathbf{H}^{\mathfrak{m},\mathfrak{n}}_{p,1}(k_w),\cdots,\mathbf{H}^{\mathfrak{m},\mathfrak{n}}_{p,q}(k_w),\cdots,\mathbf{H}^{\mathfrak{m},\mathfrak{n}}_{p,P}(k_w)],
\end{align}
where $\mathbf{H}^{\mathfrak{m},\mathfrak{n}}_{p,q}(k_w)$ is the $M\times M$ channel matrix from the $q$th $(1 \leq q \leq P)$ group elements of the $\mathfrak{n}$th UCA in UCCA of MBS to the $\mathfrak{m}$th UCA in UCCA of the $p$th SBS.

\vspace{0.0cm}
\subsection{Preprocessing for Interference Elimination in Downlink MU-OAM-MIMO System}
\vspace{-0.0cm}

As the distances and AoAs of $\mathfrak{N}$ concentric UCAs in each SBS are exactly the same, the OAM-based multi-user distance and AoA estimation and MU-OAM preprocessing methods based on a single UCA proposed in Section III and IV can be directly utilized in the UCCA-based system. Meanwhile, the number of training symbols required by the UCCA-based downlink MU-OAM-MIMO wireless backhaul system does not increase with the number of subcarriers, transmit and receive antenna elements, which is opposite in traditional MU-MIMO-OFDM channel estimation.

The effective OAM channel of the MU-OAM-MIMO system can be written as
\begin{align}\label{HOAMMIMO1}
&\bar{\mathbf{H}}_\textmd{OAM}(k_w)=\mathbf{I}_{P}\otimes\left(\mathbf{I}_{\mathfrak{N}}\otimes\mathbf{F}_{U}^{\mathrm{H}}\right)\mathbf{\bar{H}}(k_w)
\mathbf{I}_{\mathfrak{N}}\otimes\mathbf{F} \nonumber \\
&\!=\!\left[\bar{\mathbf{H}}^{\mathrm{T}}_{\textrm{OAM}_{1}}(k_w),\cdots,\bar{\mathbf{H}}^{\mathrm{T}}_{\textrm{OAM}_{p}}(k_w),\cdots,\bar{\mathbf{H}}^{\mathrm{T}}_{\textrm{OAM}_{P}}(k_w)\right]^{\mathrm{T}},
\end{align}
where $\bar{\mathbf{H}}_{\textrm{OAM}_{p}}(k_w)=\mathbf{I}_{\mathfrak{N}}\otimes\mathbf{F}_{U}^{\mathrm{H}}\mathbf{\bar{H}}_p(k_w)
\mathbf{I}_{\mathfrak{N}}\otimes\mathbf{F}$ denotes the effective OAM channel matrix from MBS to the $p$th SBS. Using the MU-OAM preprocessing method in section IV, the preprocessing matrix of MU-OAM-MIMO system can be obtained and denotes as
\begin{align} \label{MUOAMMIMOP}
\bar{\mathbf{P}}(k_w)&=\bar{\mathbf{E}}(k_w)\bar{\mathbf{G}}(k_w)\nonumber\\
&=[\bar{\mathbf{E}}_1(k_w)\bar{\mathbf{G}}_1(k_w),\cdots,\bar{\mathbf{E}}_P(k_w)\bar{\mathbf{G}}_P(k_w)].
\end{align}
where $\bar{\mathbf{E}}(k_w)=[\bar{\mathbf{E}}_1(k_w),\cdots,\bar{\mathbf{E}}_P(k_w)]$ and $\bar{\mathbf{G}}(k_w)=\textrm{diag}\{\bar{\mathbf{G}}_1(k_w),\cdots,\bar{\mathbf{G}}_P(k_w)\}$, $\bar{\mathbf{E}}_p(k_w)$ and $\bar{\mathbf{G}}_p(k_w)$ are respectively the $P\mathfrak{N}U\times \mathfrak{N}U$ inter-user interference elimination matrix and $\mathfrak{N}U\times \mathfrak{N}U$ intra-user interference elimination matrix corresponding to the $p$th SBS. The block diagram shown in Fig.\ref{Fig5} can also be applied to the downlink MU-OAM-MIMO wireless back system as long as $\mathbf{G}(k_w)$, $\mathbf{E}(k_w)$, $\mathbf{F}$, $\mathbf{H}_p(k_w)$ and $\mathbf{F}_U^\mathrm{H}$ are respectively replaced with $\bar{\mathbf{G}}(k_w)$, $\bar{\mathbf{E}}(k_w)$, $\mathbf{I}_{\mathfrak{N}}\otimes\mathbf{F}$, $\bar{\mathbf{H}}_p(k_w)$ and $\mathbf{I}_{\mathfrak{N}}\otimes\mathbf{F}_U^\mathrm{H}$.

After that, the detected OAM data symbol vector at the $p$th SBS can be written as
\begin{align} \label{xpkw'}
&\mathbf{\bar{x}}_{p}(k_w)=\mathbf{I}_{\mathfrak{N}}\otimes\mathbf{F}_{U}^{\mathrm{H}} \bigg(\mathbf{\bar{H}}_{p}(k_w)\mathbf{I}_{\mathfrak{N}}\otimes\mathbf{F}\mathbf{\bar{P}}(k_w)\mathbf{\bar{s}}(k_w)+\mathbf{\bar{z}}_p(k_w)\bigg)\nonumber\\
&=\mathbf{I}_{\mathfrak{N}}\otimes\mathbf{F}_{U}^{\mathrm{H}} \bigg(\mathbf{\bar{H}}_{p}(k_w)\mathbf{I}_{\mathfrak{N}}\otimes\mathbf{F}\sum_{q=1}^{P}\mathbf{\bar{P}}_q(k_w)\mathbf{\bar{s}}_q(k_w)+\mathbf{\bar{z}}_p(k_w)\bigg)\nonumber\\
&=\mathbf{\bar{H}}_{{\rm OAM}_p}(k_w)\mathbf{\bar{P}}_p(k_w)\mathbf{\bar{s}}_p(k_w)+\mathbf{\bar{H}}_{{\rm OAM}_p}(k_w)\mathbf{\bar{P}}'_p(k_w)\mathbf{\bar{s}}'_p(k_w)\nonumber\\
&\quad +\mathbf{\bar{z}}''_p(k_w) \nonumber\\
&=\mathbf{\bar{s}}_p(k_w)+\bm{\mathcal{\bar{I}}}^1_p(k_w)+\bm{\mathcal{\bar{I}}}^2_p(k_w)+\mathbf{\bar{z}}''_p(k_w).
\end{align}
where $\mathbf{\bar{s}}(k_w)$ $=$ $[\mathbf{\bar{s}}^T_1(k_w)$, $\mathbf{\bar{s}}^T_2(k_w)$, $\cdots$, $ \mathbf{\bar{s}}^T_{P}(k_w)]^T$, $\mathbf{\bar{s}}_p(k_w)$ $=$\\$[\mathbf{\bar{s}}^T_{p_1}(k_w)$, $\mathbf{\bar{s}}^T_{p_2}(k_w)$, $\cdots$, $ \mathbf{\bar{s}}^T_{p_\mathfrak{N}}(k_w)]^T$ is the data symbol vector transmitted to the $p$th SBS and $\mathbf{\bar{s}}_{p_\mathfrak{n}}(k_w)$ $=$ $[s^p_{\mathfrak{n}}(\ell_1,k_w),$ $s^p_{\mathfrak{n}}(\ell_2,k_w),$ $\cdots,$ $s^p_{\mathfrak{n}}(\ell_U,k_w)]^T$, $\mathbf{\bar{x}}_{p}(k_w)$ $=$ $[\mathbf{\bar{x}}^T_{p_1}(k_w)$, $\mathbf{\bar{x}}^T_{p_2}(k_w)$, $\cdots$, $ \mathbf{\bar{x}}^T_{p_\mathfrak{N}}(k_w)]^T$ is the corresponding detected data symbol vector of the $p$th SBS and $\mathbf{\bar{x}}_{p_\mathfrak{m}}(k_w)$ $=$ $[x^p_{\mathfrak{m}}(\ell_1,k_w),$ $x^p_{\mathfrak{m}}(\ell_2,k_w),$ $\cdots,$ $x^p_{\mathfrak{m}}(\ell_U,k_w)]^T$, $\mathbf{\bar{P}}'_p(k_w)$ and $\mathbf{\bar{s}}'_p(k_w)$ are defined as
\begin{align*}
\mathbf{\bar{P}}'_p(k_w)&
\!=\![\mathbf{\bar{P}}_1(k_w),  \cdots, \mathbf{\bar{P}}_{p-1}(k_w), \mathbf{\bar{P}}_{p+1}(k_w), \cdots, \mathbf{\bar{P}}_P(k_w)],
\end{align*}
\begin{align*}
\mathbf{\bar{s}}'_p(k_w)&\!=\![\mathbf{\bar{s}}^\mathrm{T}_1(k_w),  \cdots, \mathbf{\bar{s}}^\mathrm{T}_{p-1}(k_w),\mathbf{\bar{s}}^\mathrm{T}_{p+1}(k_w), \cdots, \mathbf{\bar{s}}^\mathrm{T}_{P}(k_w)],
\end{align*}
$\mathbf{\bar{P}}_p(k_w)=\bar{\mathbf{E}}_p(k_w)\bar{\mathbf{G}}_p(k_w)$, $\bm{\mathcal{\bar{I}}}^1_p(k_w)$ and $\bm{\mathcal{\bar{I}}}^2_p(k_w)$ are respectively the $\mathfrak{N}U$-dimensional inter-user interference vector and intra-user interference vector induced by the error in preprocessing matrix, and $\mathbf{\bar{z}}''_p(k_w)$ is the noise vector.

\begin{table}[t]
\small
\caption{The complexity comparison between downlink MU-OAM-MIMO and downlink MU-MIMO.}
\setlength{\belowcaptionskip}{0.0cm}   
\begin{center}
\begin{tabular}{ccc}
  \toprule
  \textbf{Scheme}                                                 &\multicolumn{2}{c}{\textbf{Complexity}}\\
  \midrule
  \multirow{2}{*}{MU-OAM-MIMO}                                &Position estimation
                                                              &$\mathcal{O}\left(\widetilde{W}\widetilde{U}\log(\widetilde{W}\widetilde{U})\right)$\\
  \specialrule{0em}{1pt}{1pt}
                                                              &Preprocessing         &$\mathcal{O}\left(WP^4\mathfrak{N}^3U^3\right)$\\
  \midrule
  \multirow{2}{*}{MU-MIMO}                                    &Channel Estimation    &$\mathcal{O}\left(WP^3\mathfrak{N}^3M^3\right)$\\
  \specialrule{0em}{1pt}{1pt}
                                                              &Preprocessing         &$\mathcal{O}\left(WP^4\mathfrak{N}^3M^3\right)$ \\
  \bottomrule
  \label{Table}
\end{tabular}
\end{center}
\vspace{-0.2cm}
\end{table}

In the presence of interferences and noise, if we define $\mathbf{\bar{R}}^p_{\rm 1}(k_w)$, $\mathbf{\bar{R}}^p_{\rm 2}(k_w)$ and $\mathbf{\bar{R}}^p_z(k_w)$ as the $\mathfrak{N}U\times \mathfrak{N}U$ covariance matrices of $\bm{\mathcal{\bar{I}}}^1_p(k_w)$, $\bm{\mathcal{\bar{I}}}^2_p(k_w)$ and $\mathbf{\bar{z}}''_p(k_w)$, i.e.,
\begin{align*}
&\mathbf{\bar{R}}^p_{\rm 1}(k_w)=\sum_{q=1,q\neq p}^{P}\!\!\big(\mathbf{\bar{H}}_{{\rm OAM}_p}(k_w)\mathbf{\bar{P}}_q(k_w)\big)
                      \mathbb{E}\left(\mathbf{\bar{s}}_q(k_w)\mathbf{\bar{s}}_q^\mathrm{H}(k_w)\right) \nonumber\\
&\qquad\qquad\qquad\qquad\!\!\big(\mathbf{\bar{H}}_{{\rm OAM}_p}(k_w)\mathbf{\bar{P}}_q(k_w)\big)^\mathrm{H}, \nonumber\\
&\mathbf{\bar{R}}^p_{\rm 2}(k_w)=\left(\mathbf{\bar{H}}_{{\rm OAM}_p}(k_w)\mathbf{\bar{P}}_p(k_w)-\mathbf{I}_{\mathfrak{N}U}\right)
\mathbb{E}\left(\mathbf{\bar{s}}_p(k_w)\mathbf{\bar{s}}_p^\mathrm{H}(k_w)\right)\nonumber\\
&\qquad\qquad\quad\!\!\left(\mathbf{\bar{H}}_{{\rm OAM}_p}(k_w)\mathbf{\bar{P}}_p(k_w)-\mathbf{I}_{\mathfrak{N}U}\right)^\mathrm{H}, \nonumber\\
&\mathbf{\bar{R}}^p_z(k_w)=\mathbb{E}\left(\mathbf{\bar{z}}''_p(k_w) (\mathbf{\bar{z}}''_p(k_w))^\mathrm{H}\right).
\end{align*}
Then, the SINR of the $p$th SBS on the $u$th mode OAM generated by the $\mathfrak{n}$th UCA of MBS at the $w$th subcarrier can be formulated as
\begin{equation*} \label{SINR}
\textrm{SINR}^{p}_{\mathfrak{n}}(\ell_u,k_w)=\frac{\mathbb{E} \left(\left|s^p_{\mathfrak{n}}(\ell_u,k_w)\right|^2\right)}{\left[\mathbf{\bar{R}}^p_{\rm 1}(k_w)\right]_{\kappa} +
\left[\mathbf{\bar{R}}^p_{\rm 2}(k_w)\right]_{\kappa}+\left[\mathbf{\bar{R}}^p_z(k_w)\right]_{\kappa}},
\end{equation*}
where $\kappa=(\mathfrak{n}-1)U+u, \mathfrak{n}=1,2,\cdots,\mathfrak{N}$. Therefore, the SE of the UCCA-based downlink MU-OAM-MIMO wireless backhaul system can be written as
\begin{align} \label{CMUU}
\bar{C}=&\bigg(1-\frac{T_t\widetilde{W}}{T_cW}\bigg) \frac{1}{W}\sum_{p=1}^P\sum_{w=1}^W\sum_{u=1}^U
 \log_2\left(1 +\textrm{SINR}^{p}_{\mathfrak{n}}(\ell_u,k_w)\right)\nonumber \\
&+\frac{1}{W}\sum_{p=1}^P\sum_{\mathfrak{n}=2}^\mathfrak{N}\sum_{w=1}^W\sum_{u=1}^U\log_2\left(1 + \textrm{SINR}^{p}_{\mathfrak{n}}(\ell_u,k_w)\right).
\end{align}
The first term in \eqref{CMUU} accounts for the SE of one UCA with training overhead for multi-user distance and AoA estimation, and the second term is the total SE of the other UCAs in UCCA at MBS.

Corresponding, the EE of the UCCA-based MU-OAM-MIMO wireless backhaul system can be written as
\begin{align}\label{EE}
\bar{\eta}_{\textrm{EE}} = \frac{B\bar{C}}{\mathcal{\bar{P}}_{\textrm{total}}} = \frac{B\bar{C}}{W\mathfrak{N}\mathcal{\bar{P}}_{\textrm{t}}/\rho + \mathcal{\bar{P}}_{\textrm{c}}},
\end{align}
where $\mathcal{\bar{P}}_{\textrm{t}}$ is the transmit power of each transmit UCA in MBS at each subcarrier, $\mathcal{\bar{P}}_{\textrm{c}}=(1+P)\mathcal{P}_{\textrm{BB}}+2\mathfrak{N}MP\mathcal{P}_{\textrm{RF}}+\mathfrak{N}MP\mathcal{P}_{\textrm{LNA}}$.

\begin{figure}[t]
\setlength{\abovecaptionskip}{-0cm}   
\setlength{\belowcaptionskip}{-0.2cm}   
\begin{center}
\includegraphics[width=8.2cm,height=6.7cm]{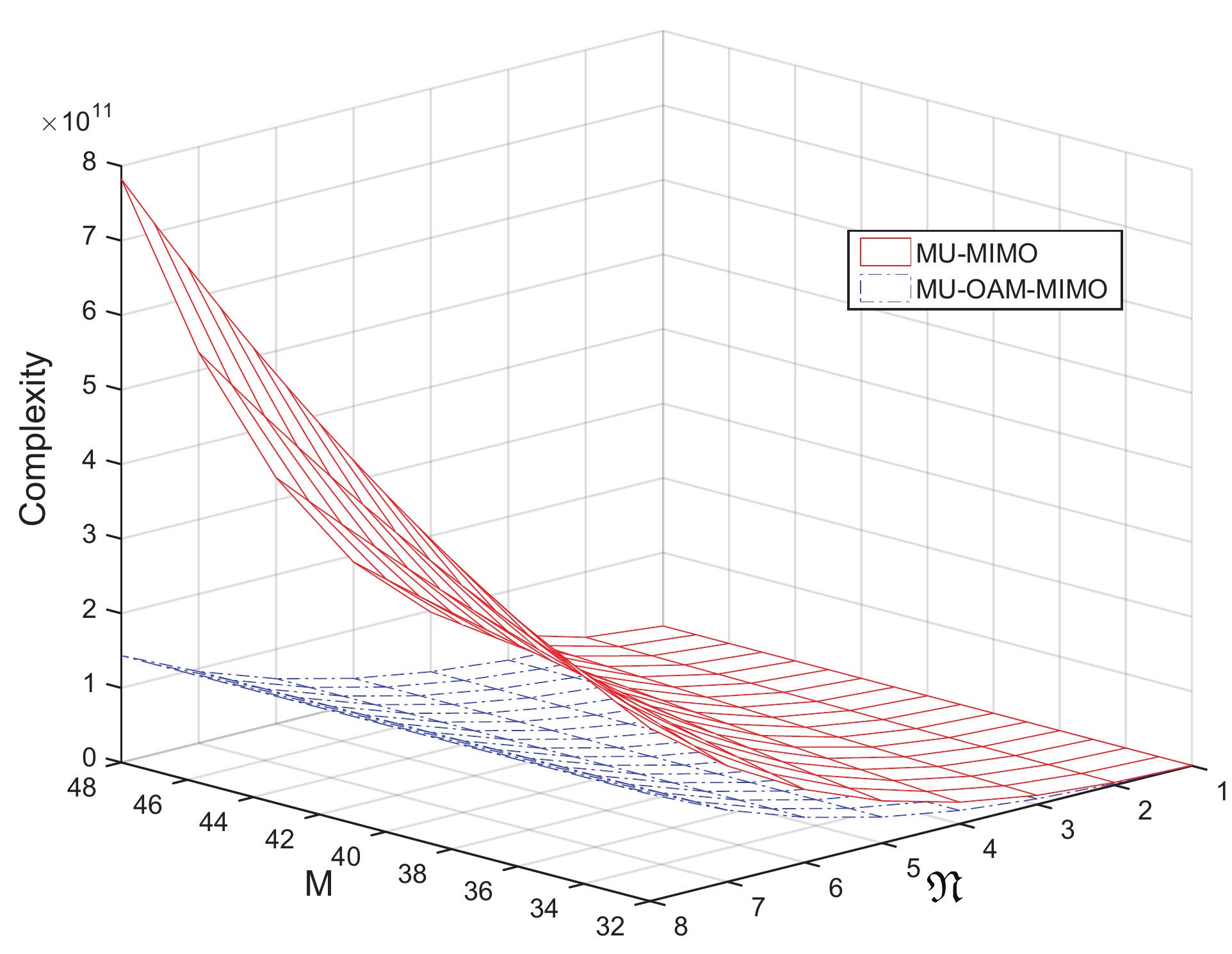}
\end{center}
\caption{The complexities of the downlink MU-OAM-MIMO wireless backhaul system and the traditional downlink MU-MIMO wireless backhaul system vs. $M$ and $\mathfrak{N}$ at $W=128, \widetilde{W}=64, U=\widetilde{U}=30$, $P=3$.}
\label{Fig6}
\end{figure}
\subsection{Performance and Complexity Discussions}
\vspace{0.0cm}
\begin{figure*}[t]
\setlength{\abovecaptionskip}{-0.0cm}   
\setlength{\belowcaptionskip}{-0.0cm}   
\begin{center}
\includegraphics[scale=0.55]{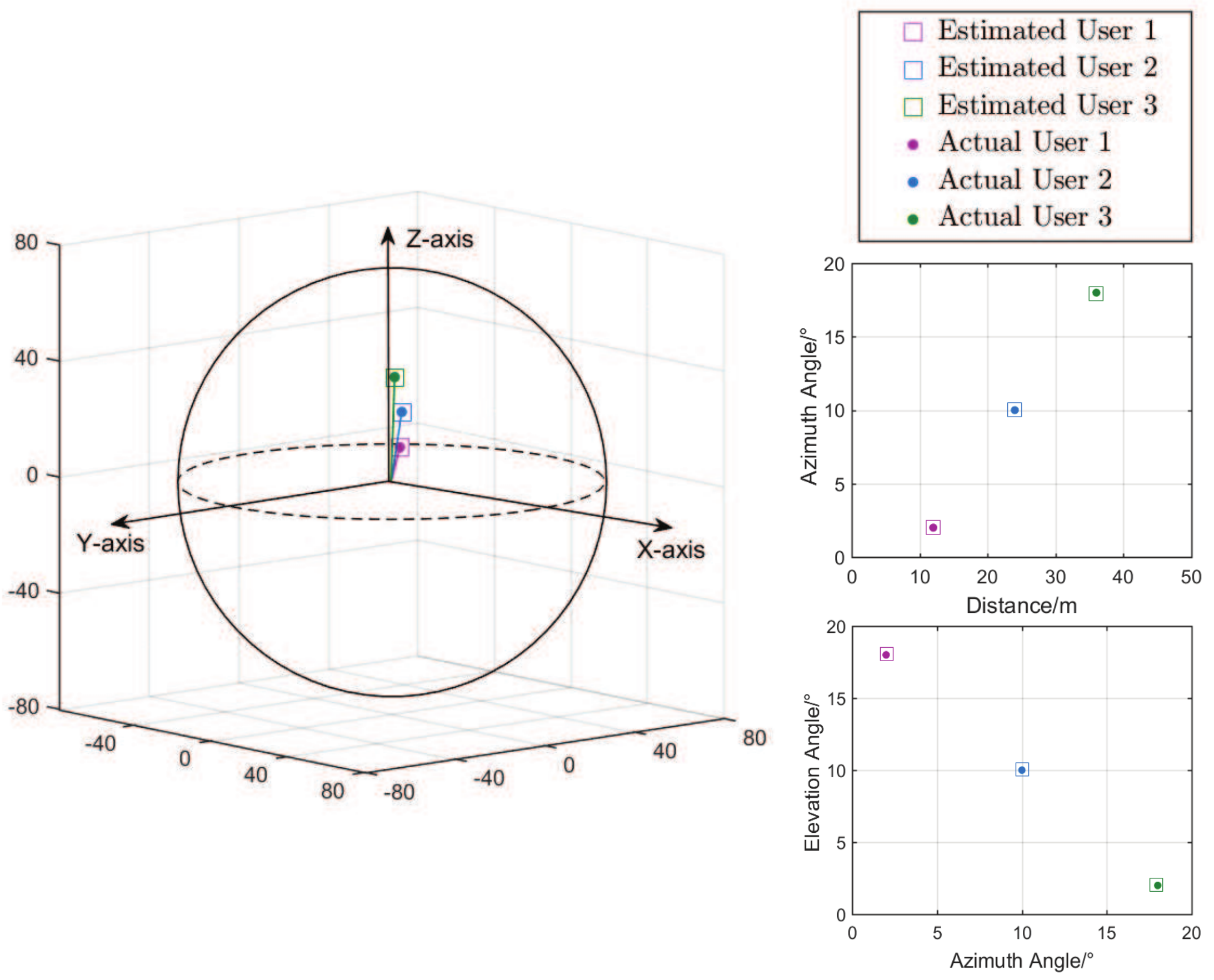}
\end{center}
\caption{The multi-user distance and AoA estimation results of the proposed method at $20dB$.}
\label{Fig7}
\end{figure*}

In this part, we mainly analyze the training overhead and computational complexity of the proposed downlink MU-OAM-MIMO wireless backhaul system with $P$ SBSs working on $W$ subcarriers and $U$ OAM modes, which is similar in structure to the traditional downlink MU-MIMO wireless backhaul system using $W$ subcarriers and $P\mathfrak{N}M\times \mathfrak{N}N$ antenna elements, where $\mathfrak{N}N=P\mathfrak{N}M$. In the traditional MU-MIMO system, to correctly recover the transmitted data symbols, channel estimation has to determine the $P\mathfrak{N}M\times \mathfrak{N}N$ channel coefficients for each subcarrier, i.e., $W\times(P\mathfrak{N}M)^2$ unknown variables in total. Thus, the required number of training symbols increases proportional to $W$, $P$, $\mathfrak{N}$ and $M$ in large-scale antenna systems. However, by taking advantage of the geometrical relationship of the UCCA-based downlink MU-OAM-MIMO wireless backhaul system, given that the number of elements and the radii of $\mathfrak{N}$ UCAs in UCCA of each SBS are known to MBS, the proposed MU-OAM-MIMO system only need determining $3P$ unknown parameters, i.e, $\{(r_p,\theta_p,\varphi_p)|p=1,2,\cdots,P\}$ to recover the transmitted data symbols. Since $\widetilde{W}$ and $\widetilde{U}$ have only effect on the accuracy of OAM-based multi-user distance and AoA estimation, when $\{(\hat{r}_p,\hat{\theta}_p,\hat{\varphi}_p)|p=1,2,\cdots,P\}$ are accurate enough, the number of training symbols required by the downlink MU-OAM-MIMO system does not need to increase with $W$, $P$, $\mathfrak{N}$ and $M$ even in large-scale antenna systems. Therefore, compared with the traditional downlink MU-MIMO wireless backhaul system, the SE of the proposed downlink MU-OAM-MIMO system can be improved due to much less training overhead.

For the downlink MU-OAM-MIMO wireless backhaul system, the complexity of the proposed OAM-based multi-user distance and AoA estimation method is determined by the complexity of 2-D FFT for \eqref{matrixX} corresponding to the values of $\widetilde{W}$, $\widetilde{U}$, the complexity of the applied MU-OAM preprocessing scheme is determined by the complexity of the SVD for $\hat{\bar{\mathbf{H}}}^{\mathrm{T}}_{\textrm{OAM}_{1}}(k_w)$ $=$ $[\bar{\mathbf{H}}^{\mathrm{T}}_{\textrm{OAM}_{1}}(k_w),$ $\cdots,$ $\bar{\mathbf{H}}^{\mathrm{T}}_{\textrm{OAM}_{p-1}}(k_w),$ $ \bar{\mathbf{H}}^{\mathrm{T}}_{\textrm{OAM}_{p+1}}(k_w),$ $\cdots\bar{\mathbf{H}}^{\mathrm{T}}_{\textrm{OAM}_{P}}(k_w)]^{\mathrm{T}}$ corresponding to the values of $W$, $P$, $\mathfrak{N}$ and $U$. The specific computational complexity of the proposed multi-user distance and AoA estimation and preprocessing for the downlink MU-OAM-MIMO wireless backhaul system is compared with that of traditional downlink MU-MIMO wireless backhaul system in the Table \ref{Table}, and the total computational complexity comparison between the downlink MU-OAM-MIMO system and the downlink MU-MIMO system is shown in the Fig.\ref{Fig6}. It can be seen from the figure that the complexity of the downlink MU-OAM-MIMO wireless backhaul system is considerably lower than the traditional downlink MU-MIMO wireless backhaul systems when $\mathfrak{N}$ and $M$ are large. For example, when $W=128$, $\widetilde{W}=64$, $U=\widetilde{U}=30$, $P=3$, $\mathfrak{N}=4$ and $M=32$, the MU-OAM-MIMO system has lower computational complexity than the traditional MU-MIMO system.

\section{Numerical Simulations and Results}

\begin{figure*}[t]
\setlength{\abovecaptionskip}{0.0cm}   
\setlength{\belowcaptionskip}{-0.0cm}   
\centering
\subfigure[]{
\includegraphics[width=5.2cm,height=4.41cm]{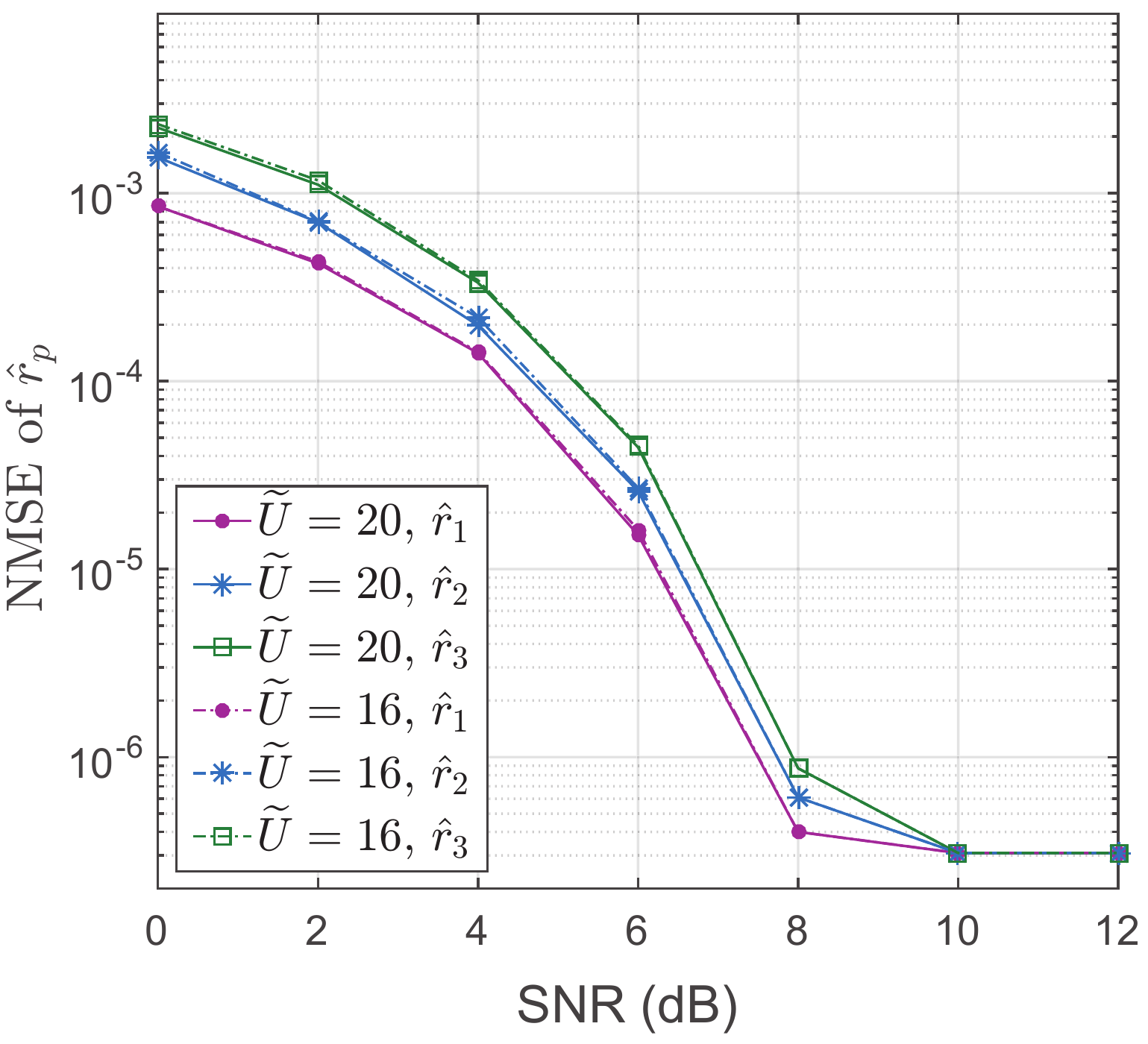}
}
\subfigure[]{
\includegraphics[width=5.21cm,height=4.41cm]{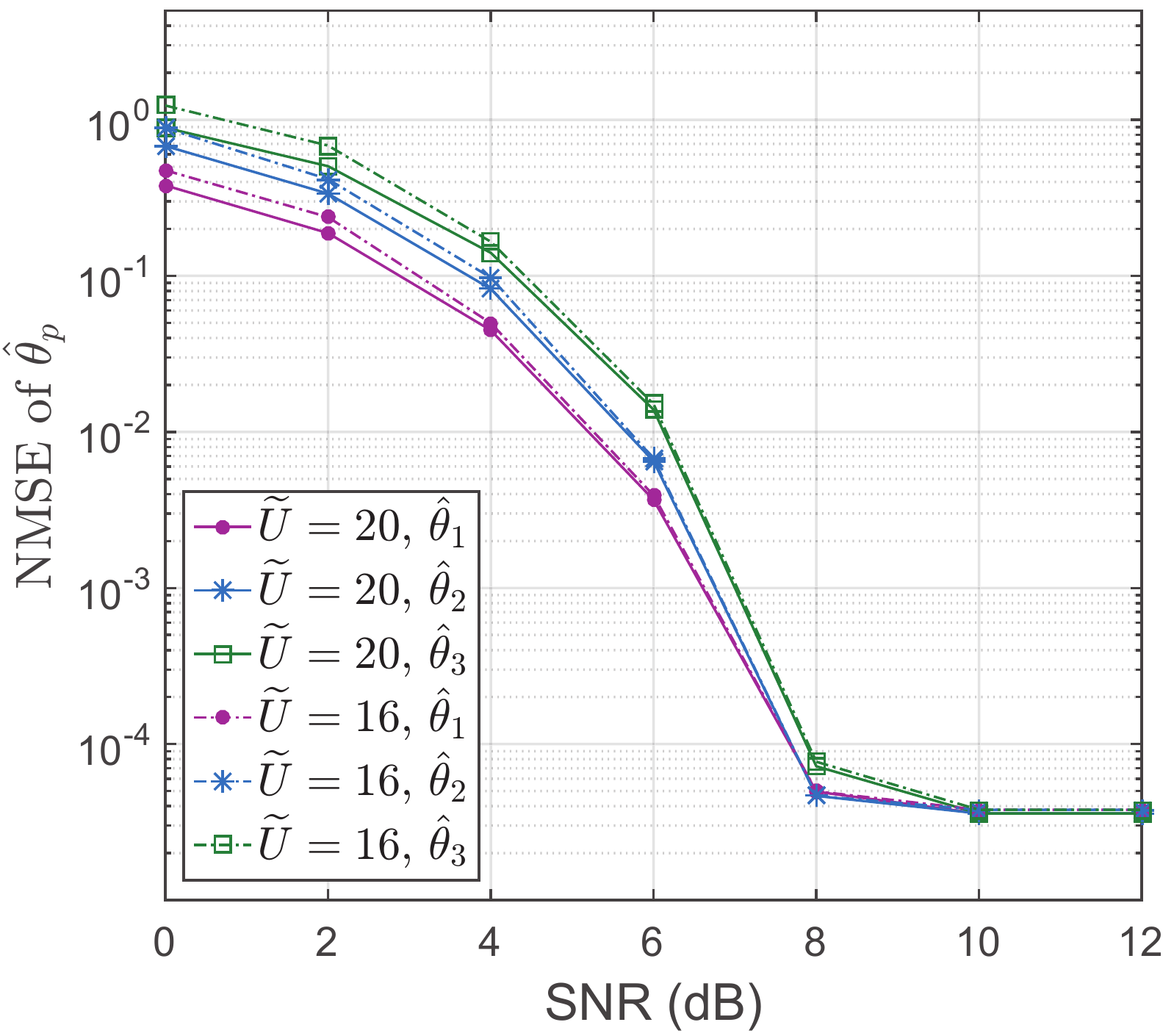}
}
\subfigure[]{
\includegraphics[width=5.2cm,height=4.4cm]{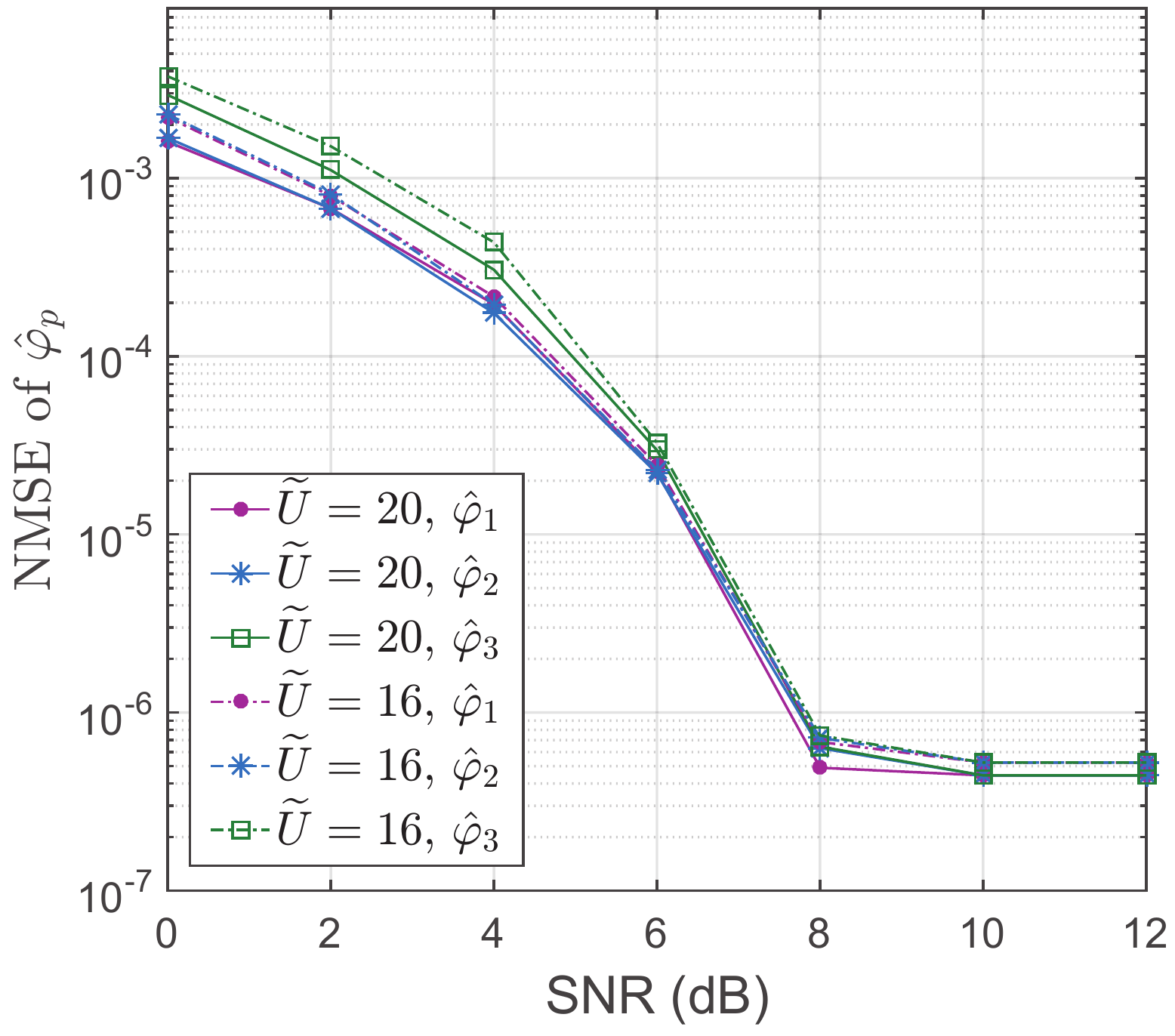}
}
\caption{The NMSEs of $\{(\hat{r}_p, \hat{\theta}_p, \hat{\varphi}_p)|p=1,2,3\}$ vs. SNR under different $\widetilde{U}$ at $\widetilde{W}=64$.}
\label{Fig8}
\end{figure*}
\begin{figure*}[t]
\setlength{\abovecaptionskip}{0.0cm}   
\setlength{\belowcaptionskip}{-0.0cm}   
\centering
\subfigure[]{
\includegraphics[width=5.0cm,height=4.39cm]{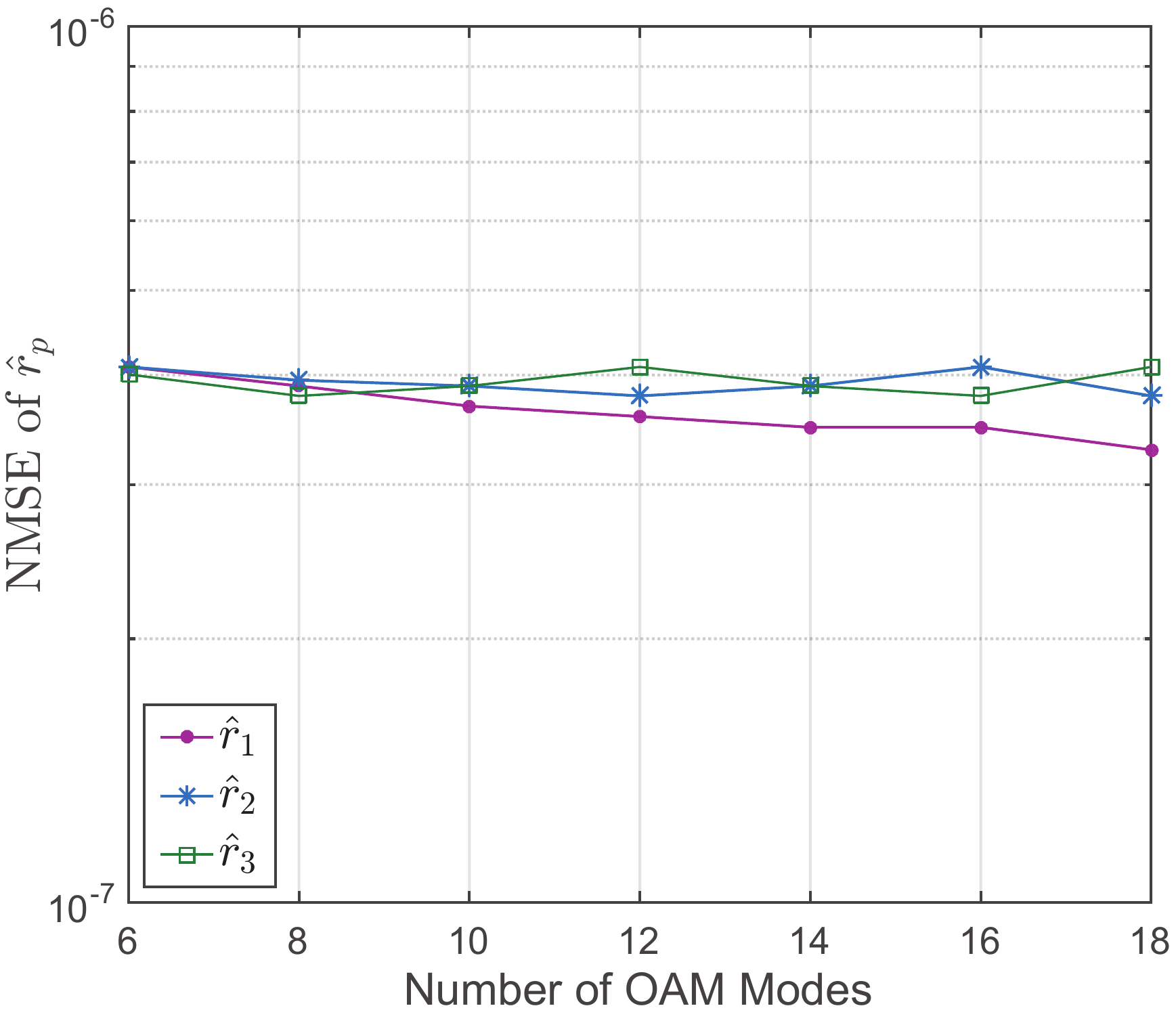}
}
\subfigure[]{
\includegraphics[width=5.05cm,height=4.3cm]{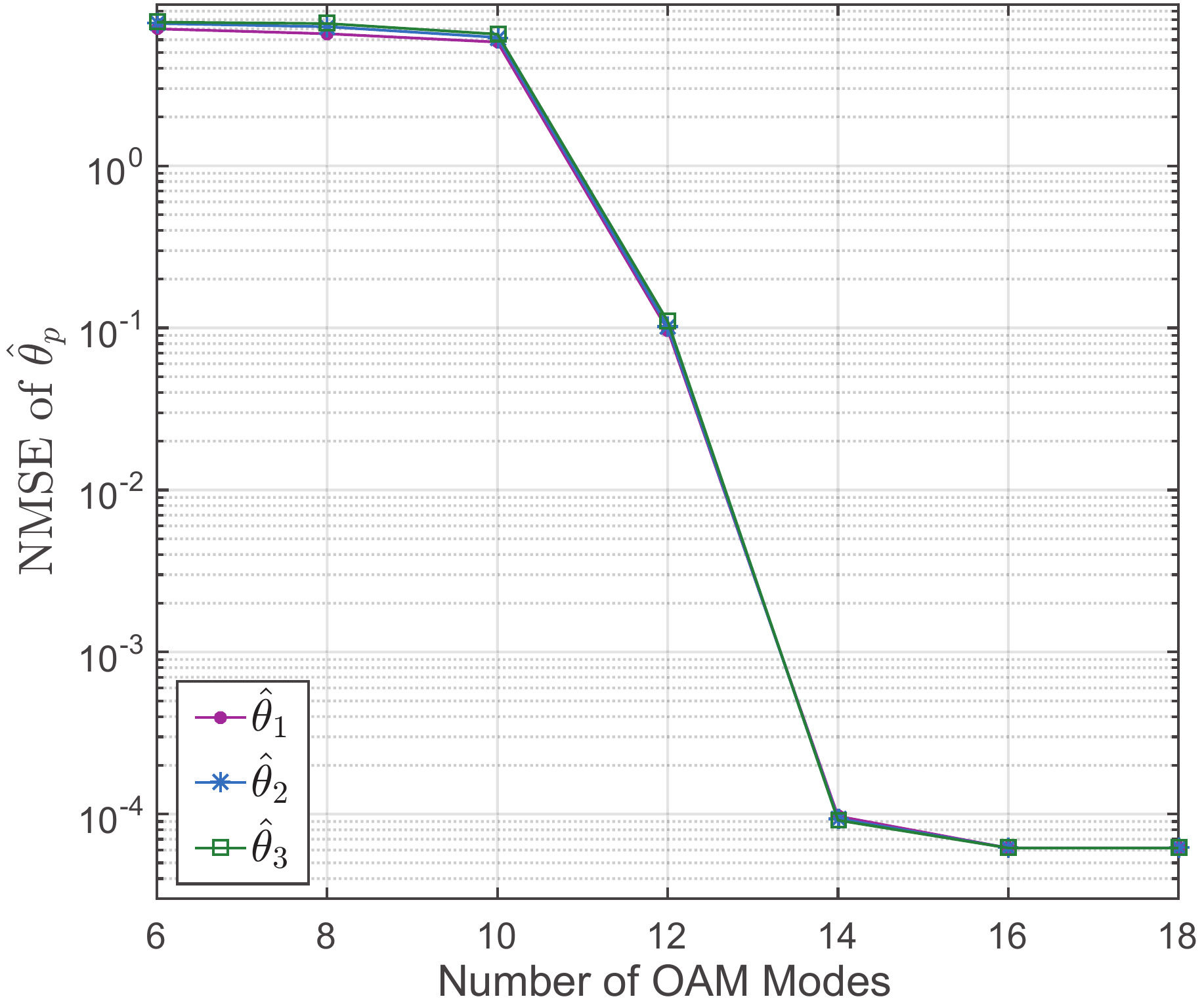}
}
\subfigure[]{
\includegraphics[width=5.0cm,height=4.32cm]{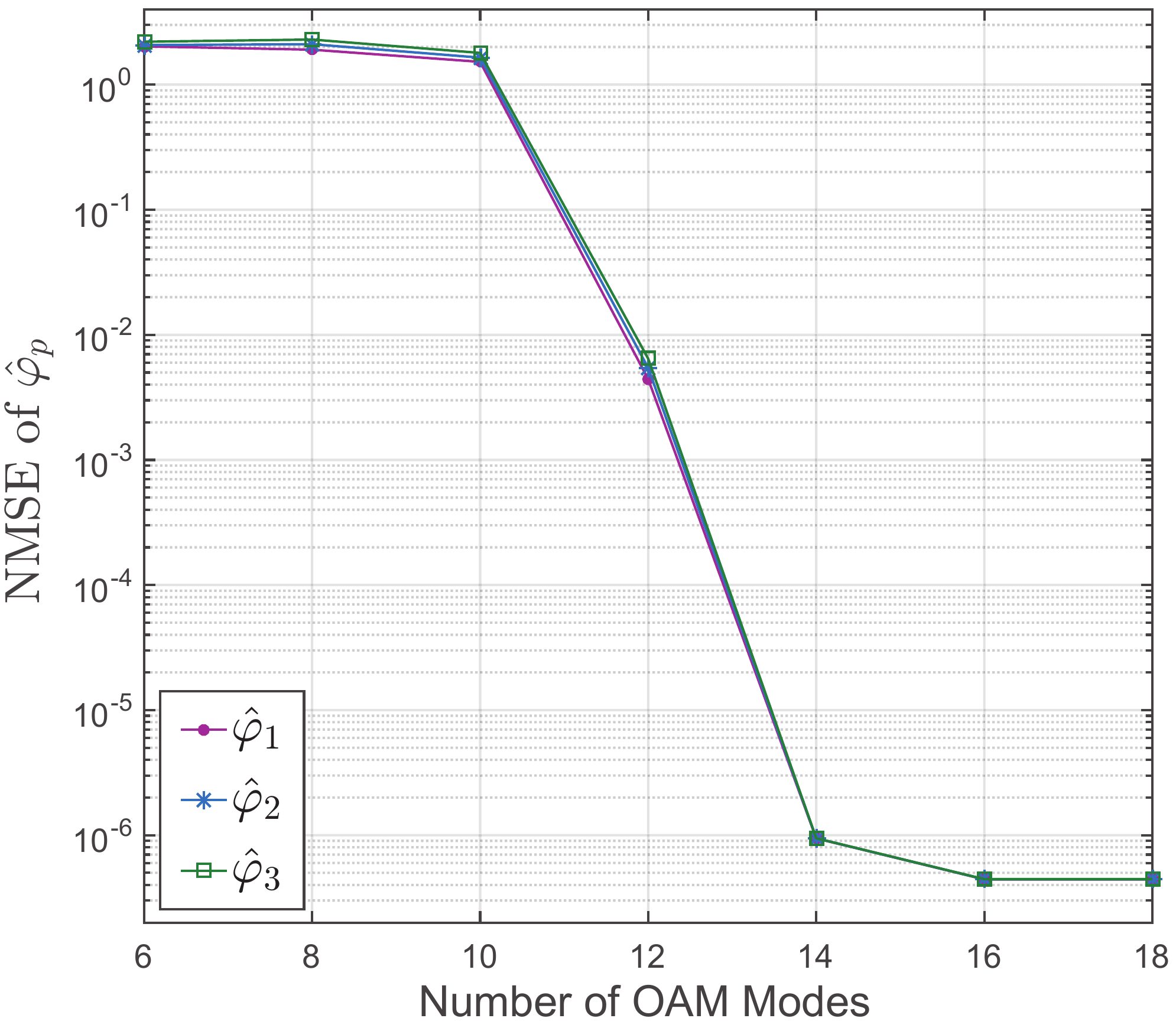}
}
\caption{The NMSEs of $\{(\hat{r}_p, \hat{\theta}_p, \hat{\varphi}_p)|p=1,2,3\}$ vs. the number of OAM modes $\widetilde{U}$ at SNR=$15$dB and $\widetilde{W}=64$.}
\label{Fig9}
\end{figure*}
\begin{figure*}[t]
\setlength{\abovecaptionskip}{0.0cm}   
\setlength{\belowcaptionskip}{-0.2cm}   
\centering
\subfigure[]{
\includegraphics[width=5.0cm,height=4.3cm]{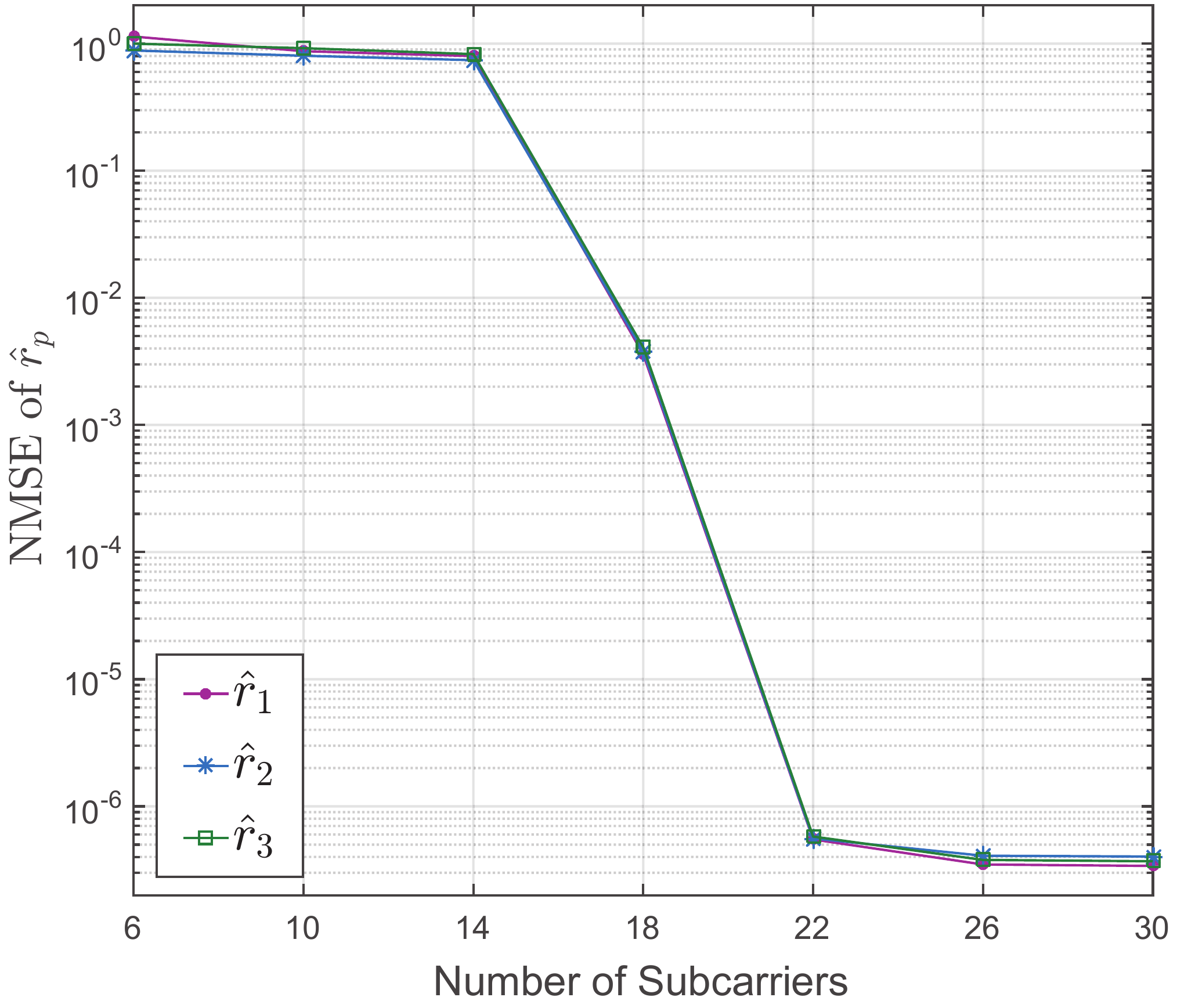}
}
\subfigure[]{
\includegraphics[width=5.0cm,height=4.3cm]{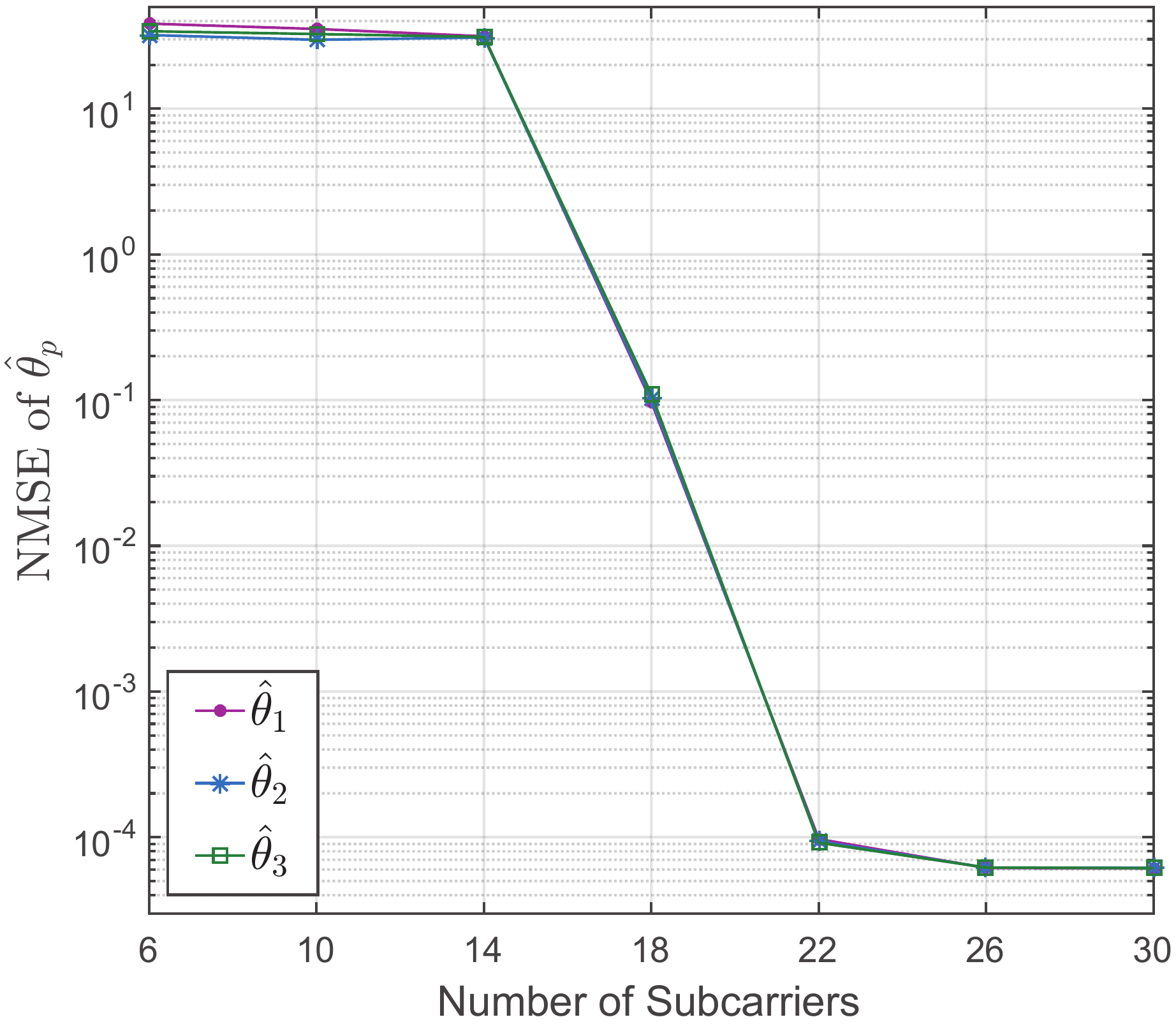}
}
\subfigure[]{
\includegraphics[width=5.0cm,height=4.4cm]{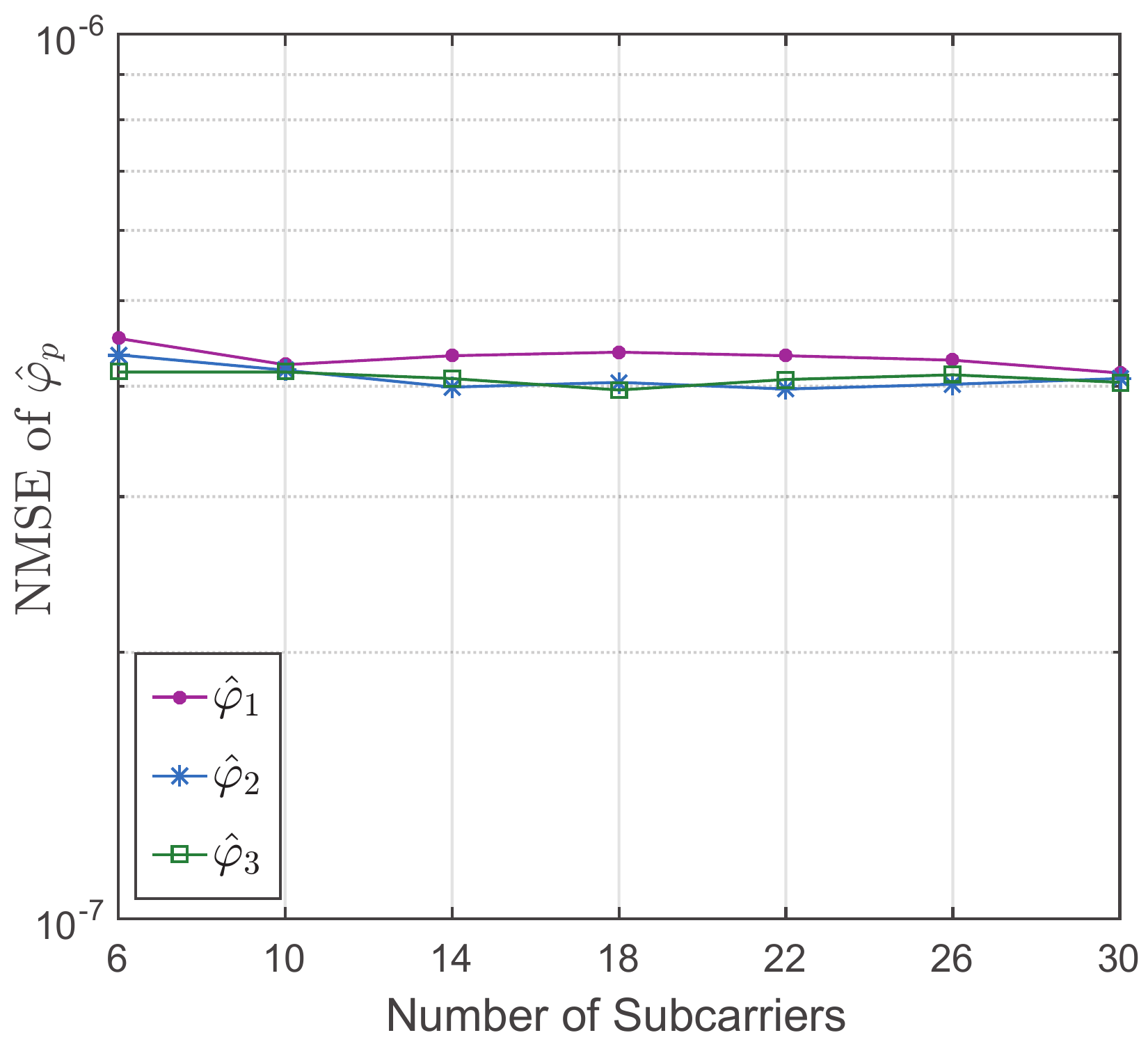}
}
\caption{The NMSEs of $\{(\hat{r}_p, \hat{\theta}_p, \hat{\varphi}_p)|p=1,2,3\}$ vs. the number of subcarriers $\widetilde{W}$ at SNR=$15$dB and $\widetilde{U}=20$.}
\label{Fig10}
\end{figure*}

In this section, we show the performances of the proposed methods by numerical simulations. We first verify the proposed OAM-based multi-user distance and AoA estimation method, and then compare the BER performances of the applied MU-OAM preprocessing scheme with those in the ideal downlink MU-OAM channel. Besides, the SEs of the UCA-based downlink MU-OAM wireless backhaul system, the UCCA-based downlink MU-OAM-MIMO wireless backhaul system and the traditional downlink MU-MIMO wireless backhaul system are compared under different multiplexed OAM modes. At last, the EEs of the UCA-based MU-OAM system and the UCCA-based MU-OAM-MIMO system are illustrated and compared with that of the traditional MU-MIMO system. Unless otherwise stated, the signal to noise ratios (SNRs) in all the figures are defined as the ratio of the received signal power versus the noise power.

In Fig.\ref{Fig7}, we choose $\widetilde{W}$ $=$ $64$ subcarriers at $f$ $=$ $9\rm GHz$ and ${\widetilde{U}} = 20$ OAM modes with $\ell_1,$ $\ell_2,$ $\cdots,$ $\ell_{20}$ $=$ $-10,$ $-9,$ $\cdots,$ $+9$, $M=21$, $P=3$, $R_t$ $ =$ $30\lambda$, $R_{r_1}$ $=$ $R_{r_2}$ $=$ $R_{r_3}$ $=$ $15\lambda$, $\lambda=c/f$, $(r_1, \theta_1, \varphi_1) = (12\textrm{m}, 18^{\circ}, 2^{\circ})$, $(r_2, \theta_2, \varphi_2) = (24\textrm{m}, 10^{\circ}, 10^{\circ})$ and $(r_3, \theta_3, \varphi_3) = (36\textrm{m}, 2^{\circ}, 18^{\circ})$. Then, by using the proposed OAM-based multi-user distance and AoA estimation method, the estimated locations of $P$ SBSs in the downlink MU-OAM wireless backhaul system are shown in Fig.\ref{Fig7}. It is obvious to see from the figure that the estimated locations of $P$ SBSs are very close to the actual locations at $20dB$, which proves the validity of the proposed method.

Fig.\ref{Fig8} illustrates the normalized mean-squared errors (NMSEs) of $\{(\hat{r}_p, \hat{\theta}_p, \hat{\varphi}_p)|p=1,2,3\}$ versus SNR under different $\widetilde{U}$. The NMSE is defined as $\mathbb{E}\{{(\hat{x}-x)}^2/x^2\}$, where $\hat{x}$ denotes the estimate of $x$. It can be seen from Fig.\ref{Fig8} that the estimation errors of $\{(\hat{r}_p, \hat{\theta}_p, \hat{\varphi}_p)|p=1,2,3\}$ are relatively small even at low and moderate SNRs. Meanwhile, as ${\widetilde{U}}$ increase the NMSEs of $\hat{\varphi}_p$ and $\hat{\theta}_p$ slightly decreases at low SNRs. Besides, it is obvious to see from the figure that compared with $\hat{r}_p$ and $\hat{\varphi}_p$, $\hat{\theta}_p$ has a larger estimation error due to the accumulation of estimation errors.

Fig.\ref{Fig9} and Fig.\ref{Fig10} show the NMSEs of $\{(\hat{r}_p, \hat{\theta}_p, \hat{\varphi}_p)|p=1,2,3\}$ versus the number of OAM modes $\widetilde{U}$ and the number of subcarriers $\widetilde{W}$. It can be seen from the two figures that when the SNR is enough high, the estimation accuracy of $\hat{r}_p$ is mainly dependent on the the number of subcarriers $\widetilde{W}$, independent on the number of OAM modes $\widetilde{U}$. Besides, the estimation accuracy of $\hat{\varphi}_p$ is mainly dependent on the number of OAM modes $\widetilde{U}$, independent on the the number of subcarriers $\widetilde{W}$. These results are consistent with the characteristics of 2-D FFT algorithm. In contrast to $\hat{r}_p$ and $\hat{\varphi}_p$, the estimation accuracy of $\hat{\theta}_p$ is simultaneously restricted by the number of $\widetilde{U}$ and $\widetilde{W}$ due to the estimation process \eqref{est_pro}.

\begin{figure}[t]
\setlength{\abovecaptionskip}{-0.1cm}   
\setlength{\belowcaptionskip}{-0.1cm}   
\begin{center}
\includegraphics[width=8.1cm,height=7.5cm]{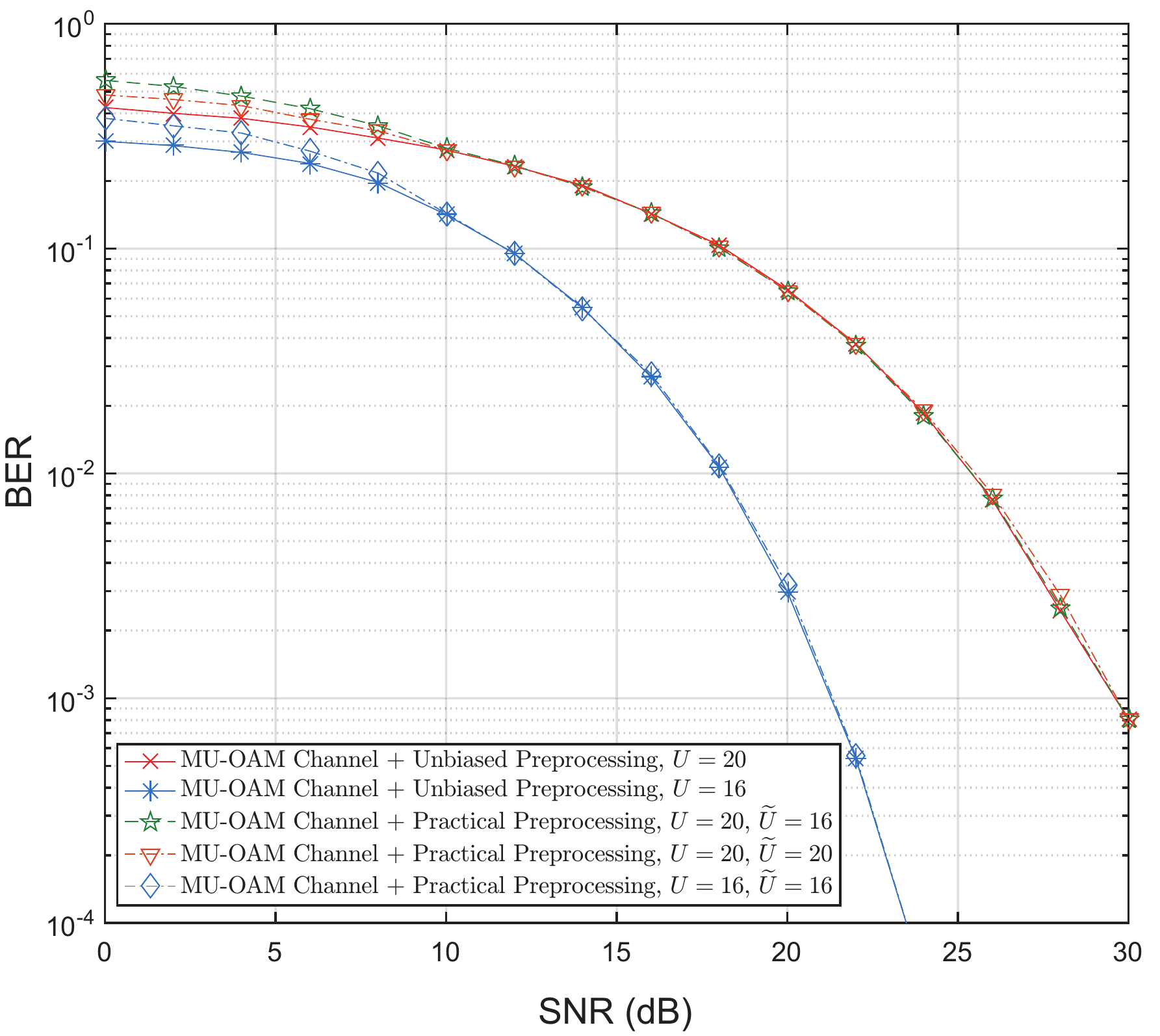}
\end{center}
\caption{The BER comparison of MU-OAM Preprocessing.}
\label{Fig11}
\end{figure}

After obtaining $\{(\hat{r}_p, \hat{\theta}_p, \hat{\varphi}_p)|p=1,2,3\}$, Fig.\ref{Fig11} compares the BERs of the applied MU-OAM preprocessing scheme with those in the ideal downlink MU-OAM channel under QPSK modulation. The ideal downlink MU-OAM channel is the downlink MU-OAM channel that is unbiased preprocessed by the actual distances and AoAs of $P$ SBSs. When $U=20$, the BERs of the applied MU-OAM preprocessing scheme are very close to those in the ideal downlink MU-OAM channel at high SNR. Meanwhile, when using fewer OAM modes (i.e.,$\widetilde{U}=16$ with $\ell_1,\ell_2,\cdots,\ell_{16} = -8,-7,\cdots,+7$) in the OAM-based multi-user distance and AoA estimation, the BER of the applied MU-OAM preprocessing scheme slightly increases at low SNR due to slightly higher estimation errors. Besides, comparing $U=16$ with $U=20$, the BERs of both the practical MU-OAM channel and the ideal MU-OAM channel all become much better because of abandoning the high-order OAM modes with smaller gains.

\begin{figure}[t]
\setlength{\abovecaptionskip}{0.0cm}   
\setlength{\belowcaptionskip}{-0.3cm}   
\begin{center}
\includegraphics[width=8.0cm,height=7.20cm]{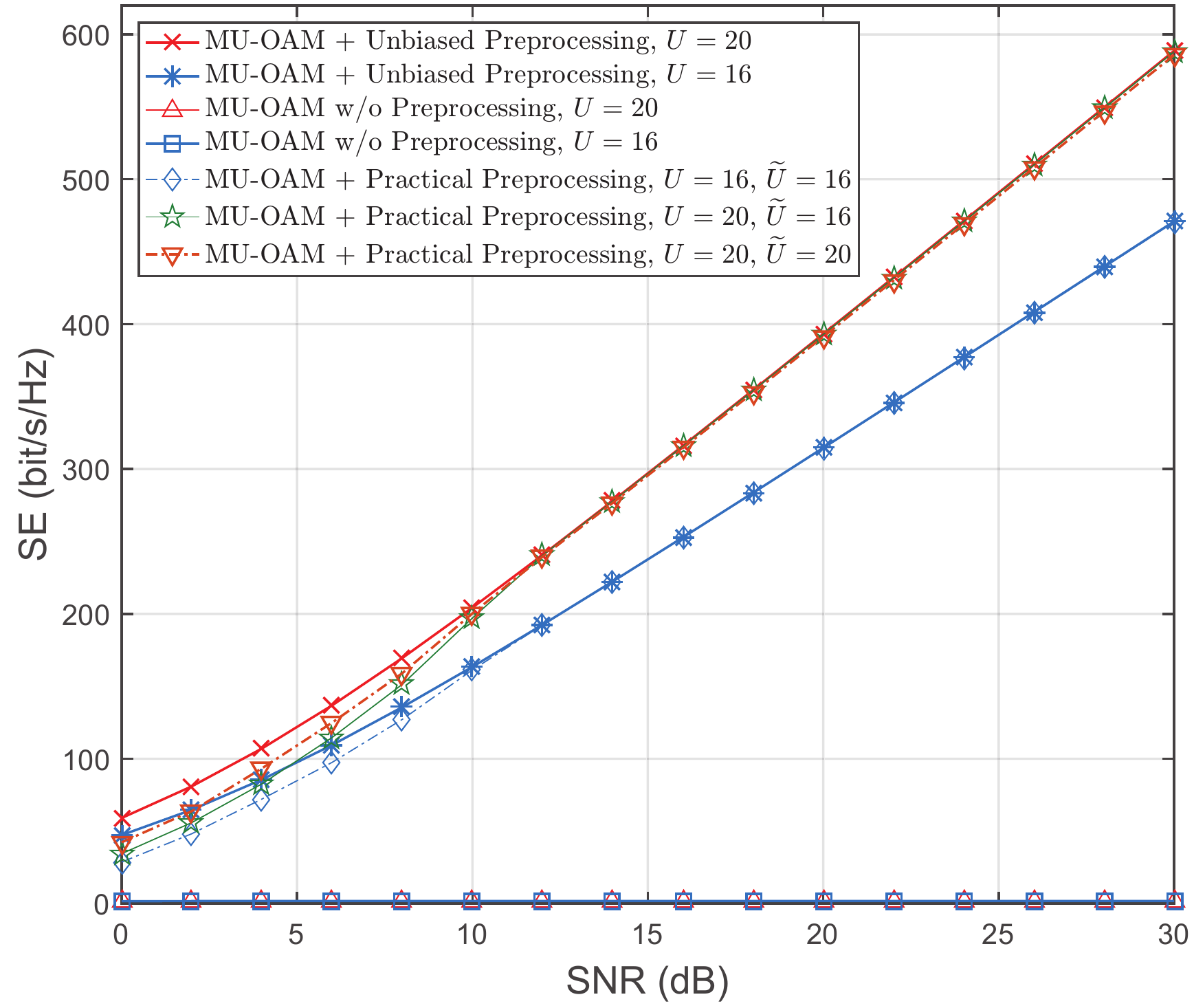}
\end{center}
\caption{The SEs of the UCA-based downlink MU-OAM wireless backhaul system.}
\label{Fig12}
\end{figure}
\begin{figure}[t]
\setlength{\abovecaptionskip}{0.05cm}   
\setlength{\belowcaptionskip}{-0.0cm}   
\begin{center}
\includegraphics[width=8.1cm,height=7.35cm]{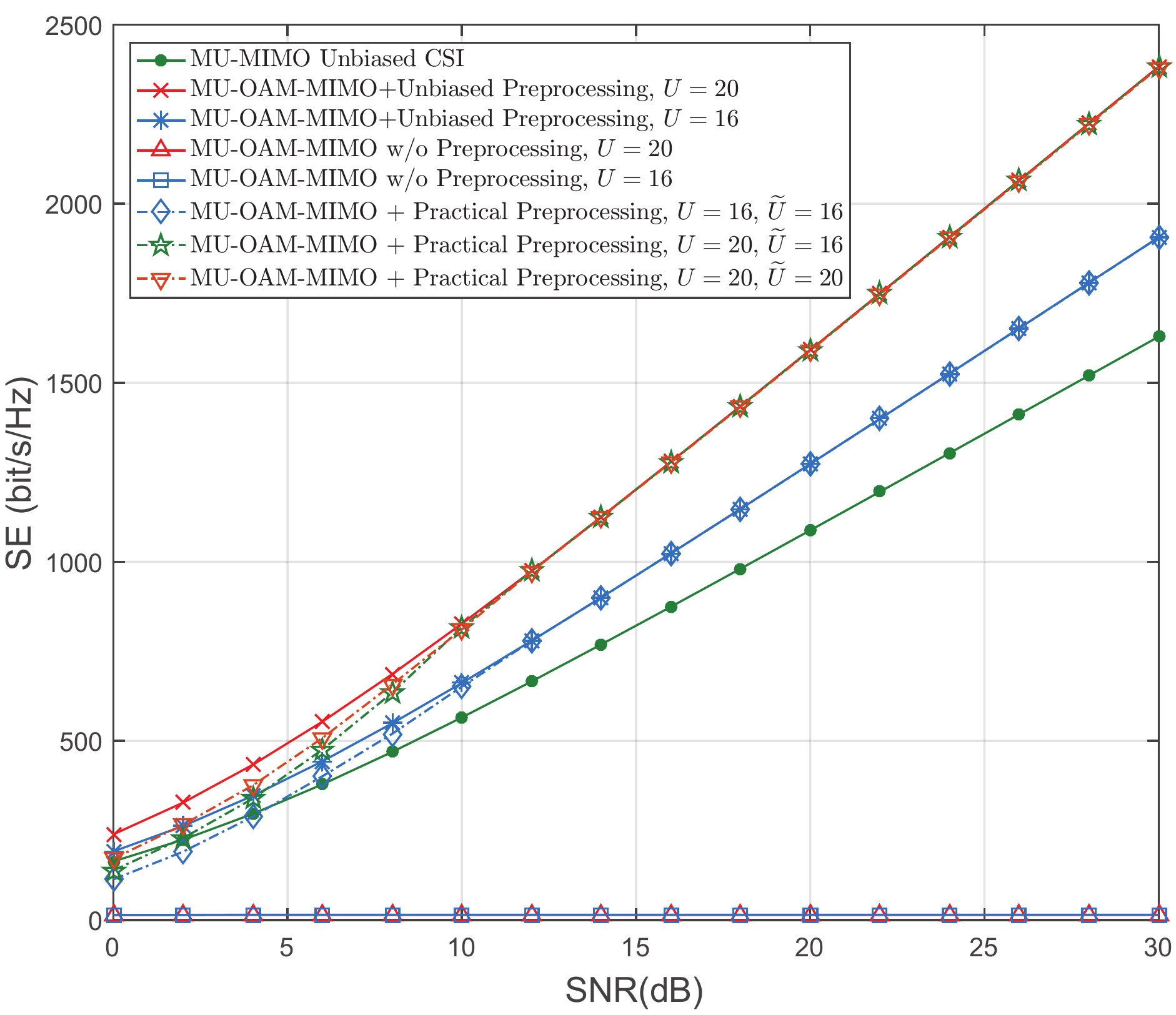}
\end{center}
\caption{The SEs of the UCCA-based downlink MU-OAM-MIMO wireless backhaul system and the traditional downlink MU-MIMO wireless backhaul system.}
\label{Fig13}
\end{figure}
\begin{figure}[t]
\setlength{\abovecaptionskip}{0.2cm}   
\setlength{\belowcaptionskip}{-0.0cm}   
\centering
\begin{minipage}{4.3cm}
\includegraphics[width=4.2cm,height=7.25cm]{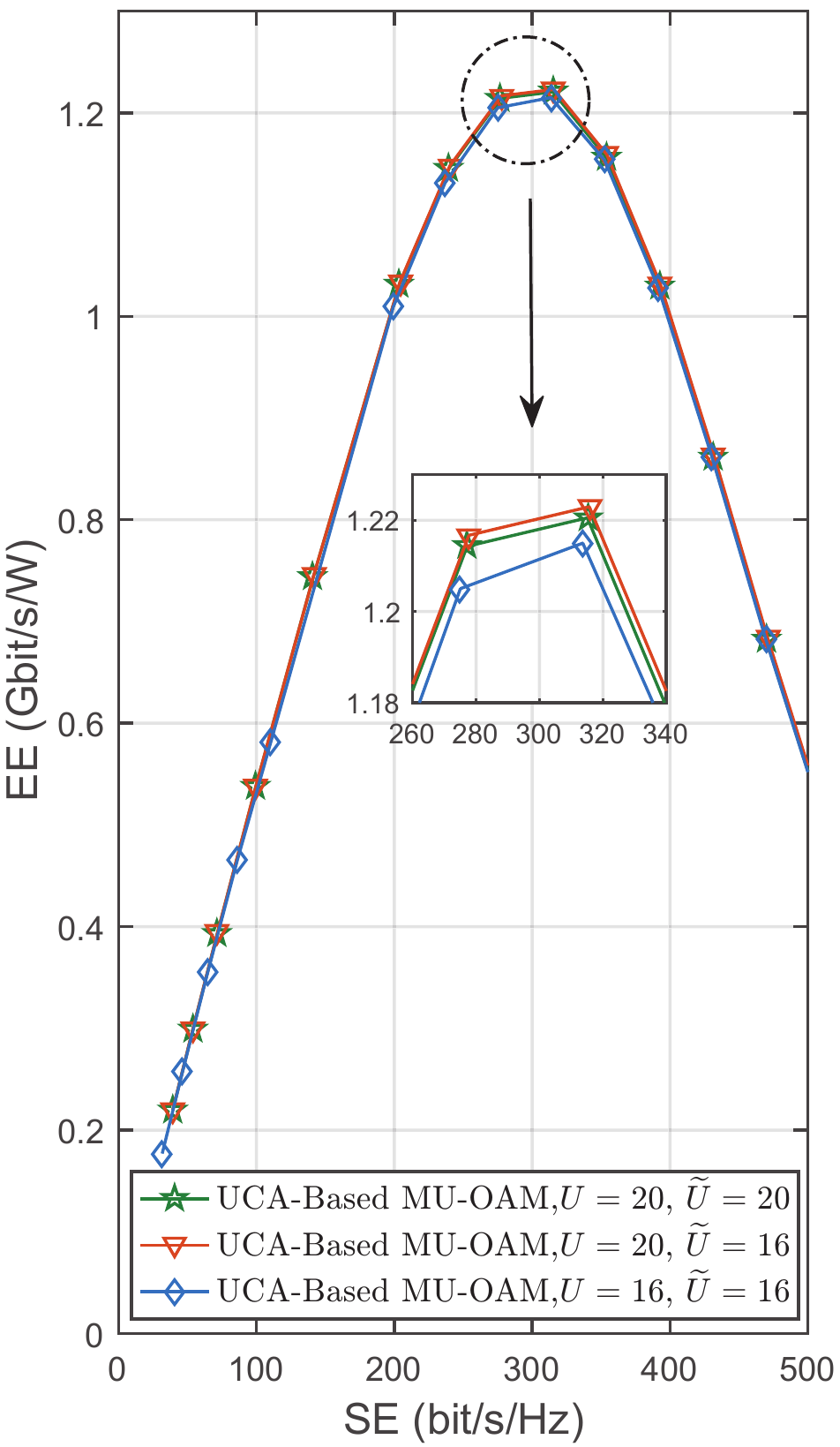}
\caption*{\footnotesize(a)}
\end{minipage}
\begin{minipage}{4.3cm}%
\includegraphics[width=4.3cm,height=7.20cm]{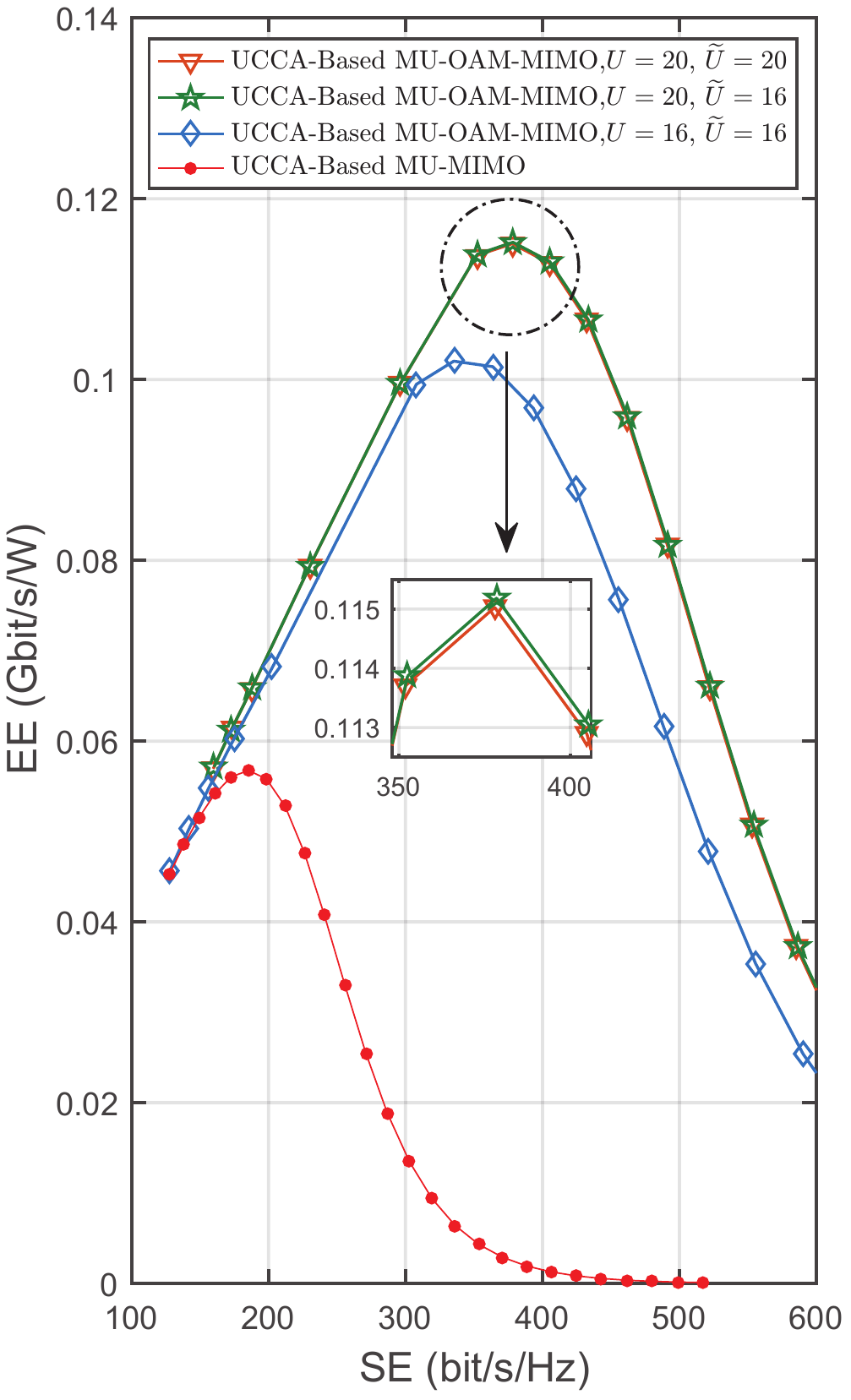}
\caption*{\footnotesize(b)}
\end{minipage}
\caption{The EE-SE relationships. (a) UCA-based downlink MU-OAM wireless backhaul system. (b) UCCA-based downlink MU-OAM-MIMO wireless backhaul system and the traditional UCCA-based downlink MU-MIMO wireless backhaul system with $\mathfrak{N}=4$.}
\label{Fig14}
\end{figure}

In Fig.\ref{Fig12}, $T_c$ is assumed to be $512$ OFDM symbols, $W=128$. Then, the SEs of the UCA-based downlink MU-OAM wireless backhaul system under different values of $U$ and $\widetilde{U}$ are shown in Fig.\ref{Fig12}. It can be seen from the figure that in contrast to the MU-OAM system without preprocessing, the SEs of the MU-OAM system with preprocessing by the OAM-based multi-user distance and AoA estimation are greatly improved, which get close to the SEs of the ideal downlink MU-OAM system at high SNR. Meanwhile, when $20$ OAM modes rather than $16$ OAM modes being used in the multi-user distance and AoA estimation, the SE of the downlink MU-OAM system increases slightly at low SNR due to lower estimation error. Furthermore, comparing $U=20$ with $U=16$, the SEs of both the practical downlink MU-OAM system and the ideal downlink MU-OAM system all become higher due to that more OAM modes are multiplexed to transmit data.

In Fig.\ref{Fig13}, $T_c$ is also assumed to be $512$ OFDM symbols, $M=21$, $P=3$, $\mathfrak{N}=4$, $W=128$ and $\widetilde{W}=64$. The SEs of the UCCA-based downlink MU-OAM-MIMO wireless backhaul system and the traditional downlink MU-MIMO wireless backhaul system are compared in Fig.\ref{Fig13}. We assume that the traditional channel estimation of the MU-MIMO system requires $P\mathfrak{N}M=252$ training symbols per subcarrier or subchannel. From the figure one can get the same conclusion as in Fig.\ref{Fig12}, which verifies that the proposed multi-user distance and AoA estimation method and the MU-OAM preprocessing scheme can be applied to the UCCA-based downlink MU-OAM-MIMO wireless backhaul system. More importantly, it is obvious that the SEs of the UCCA-based downlink MU-OAM-MIMO system are about 30\% higher than the SE of the downlink MU-MIMO system with traditional channel estimation due to the greatly reduced number of training symbols.

In Fig.\ref{Fig14}, we choose the system available bandwidth $B$ $=$ $190$MHz with $W$ $=$ $128$ subcarriers, subcarrier interval $\Delta f$ $=$ $1.48$MHz, $\mathcal{P}_{\textrm{BB}}$ $=$ $200$mW, $\mathcal{P}_{\textrm{RF}}$ $=$ $250$mW, $\mathcal{P}_{\textrm{LNA}}$ $=$ $20$mW and $\rho$ $=$ $0.35$ \cite{Chen2020Spectral,Lin2016Energy,Rial2016Hybrid}. Besides, we assume that all systems in Fig.\ref{Fig14} have the same total transmit power. Then, the EE-SE relationship of the UCA-based downlink MU-OAM wireless backhaul system under different values of $U$ and $\widetilde{U}$ is shown in Fig.\ref{Fig14} (a), and the EE of the UCCA-based MU-OAM-MIMO system is illustrated and compared with that of the traditional MU-MIMO system in Fig.\ref{Fig14} (b). It can be seen from Fig.\ref{Fig14} (a) that the EE of the proposed MU-OAM system does not always increase with the SE, and there are tradeoffs between the EE and the SE. Meanwhile, under the same transmit power, the EE of the MU-OAM system with $U=20$ is slightly higher than that with $U=16$ owing to slightly higher capacity. Besides, when using fewer OAM modes (i.e.,$\widetilde{U}=16$) in the training stage, the EE of the MU-OAM system slightly increases due to less training overhead. From Fig.\ref{Fig14} (b) one can get the same conclusion as in Fig.\ref{Fig14} (a). Moreover, under the same transmit power, the EE of the MU-OAM-MIMO system is lower than that of the MU-OAM system due to larger circuit power consumption. More importantly, it is obvious that the EE of the MU-OAM-MIMO system is almost twice that of the traditional MU-MIMO system due to the greatly reduced number of training symbols.

\section{Conclusions}

In this paper, we propose a UCA-based downlink MU-OAM wireless backhaul system, which enables joint spatial division and coaxial multiplexing. Then, we design an overall communication scheme for the downlink MU-OAM wireless backhaul system including the OAM-based multi-user distance and AoA estimation and the preprocessing of MU-OAM signals. The proposed multi-user distance and AoA estimation method is shown being able to simultaneously estimate the position of each SBS with a flexible number of training symbols. Thereafter, we apply a MU-OAM preprocessing scheme for the downlink MU-OAM wireless backhaul system, which can effectively eliminate both the co-mode and inter-mode interferences in the downlink MU-OAM channel. Since both interferences are eliminated through preprocessing at MBS, the complexity of SBSs in proposed system can be reduced. At last, the proposed methods are extended to the UCCA-based downlink MU-OAM-MIMO wireless backhaul system, which exhibits higher SE and EE than the large-scale MU-MIMO wireless backhaul system with traditional channel estimation.

\section*{Acknowledgment}
The authors would like to thank the editor and the anonymous reviewers for their careful reading and valuable suggestions that helped to improve the quality of this manuscript.

%
\bibliographystyle{IEEEtran}
\bibliography{IEEEabrv,mu_OAM}
\begin{IEEEbiography}[{\includegraphics[width=1in,height=1.25in,clip,keepaspectratio]{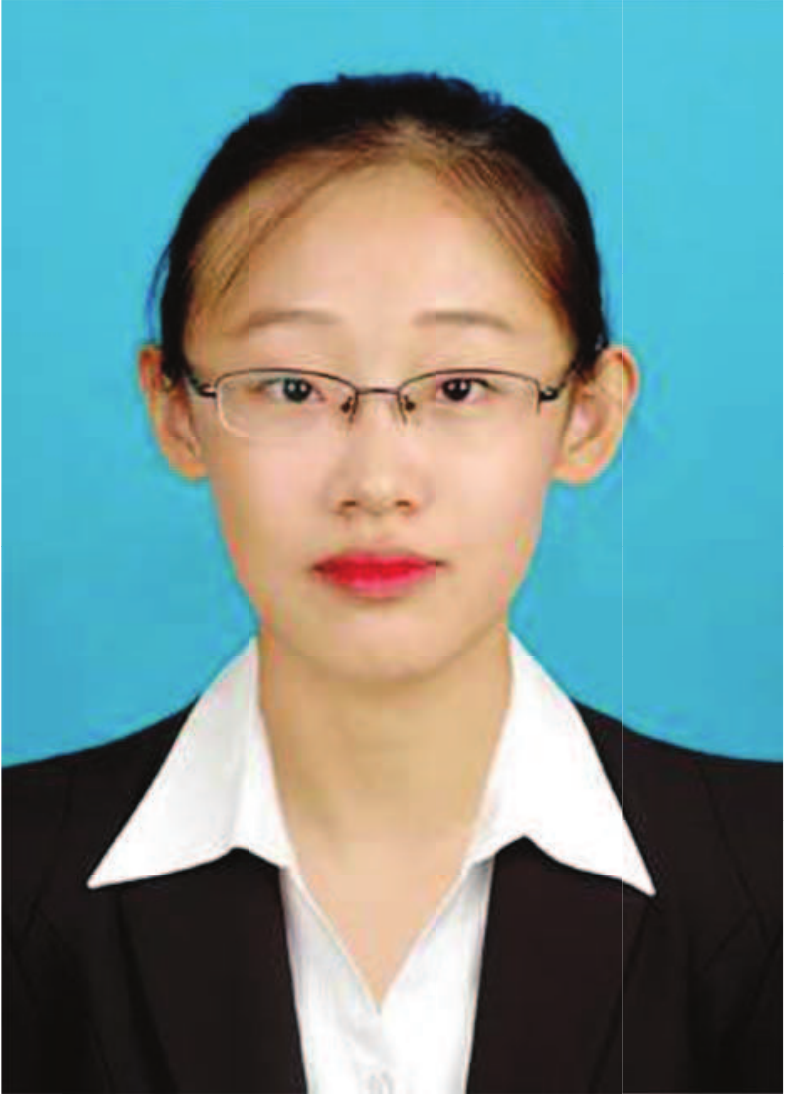}}]{Wen-Xuan Long}(Graduate Student Member, IEEE)
received the B.S. degree (with Highest Hons.) in Rail Transit Signal and Control from Dalian Jiaotong University, Dalian, China in 2017. She is currently pursuing a double Ph.D. degree in Communications and Information Systems at Xidian University, China and University of Pisa, Italy. Her research interests include broadband wireless communication systems and array signal processing.
\end{IEEEbiography}

\begin{IEEEbiography}[{\includegraphics[width=1in,height=1.25in,clip,keepaspectratio]{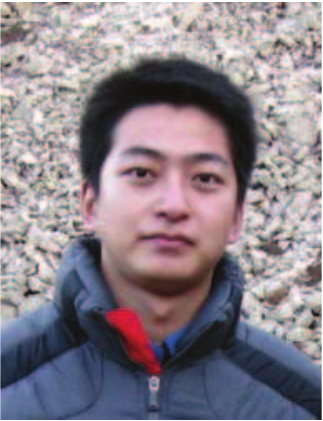}}]{Rui Chen}(Member, IEEE)
received the B.S., M.S. and Ph.D. degrees in Communications and Information Systems from Xidian University, Xi'an, China, in 2005, 2007 and 2011, respectively. From 2014 to 2015, he was a visiting scholar at Columbia University in the City of New York. He is currently an associate professor and Ph.D. supervisor in the school of Telecommunications Engineering at Xidian University. He has published about 50 papers in international journals and conferences and held 10 patents. He is an Associate Editor for International Journal of Electronics, Communications, and Measurement Engineering (IGI Global). His research interests include broadband wireless communication systems, array signal processing and intelligent transportation systems.
\end{IEEEbiography}

\begin{IEEEbiography}[{\includegraphics[width=1in,height=1.25in,clip,keepaspectratio]{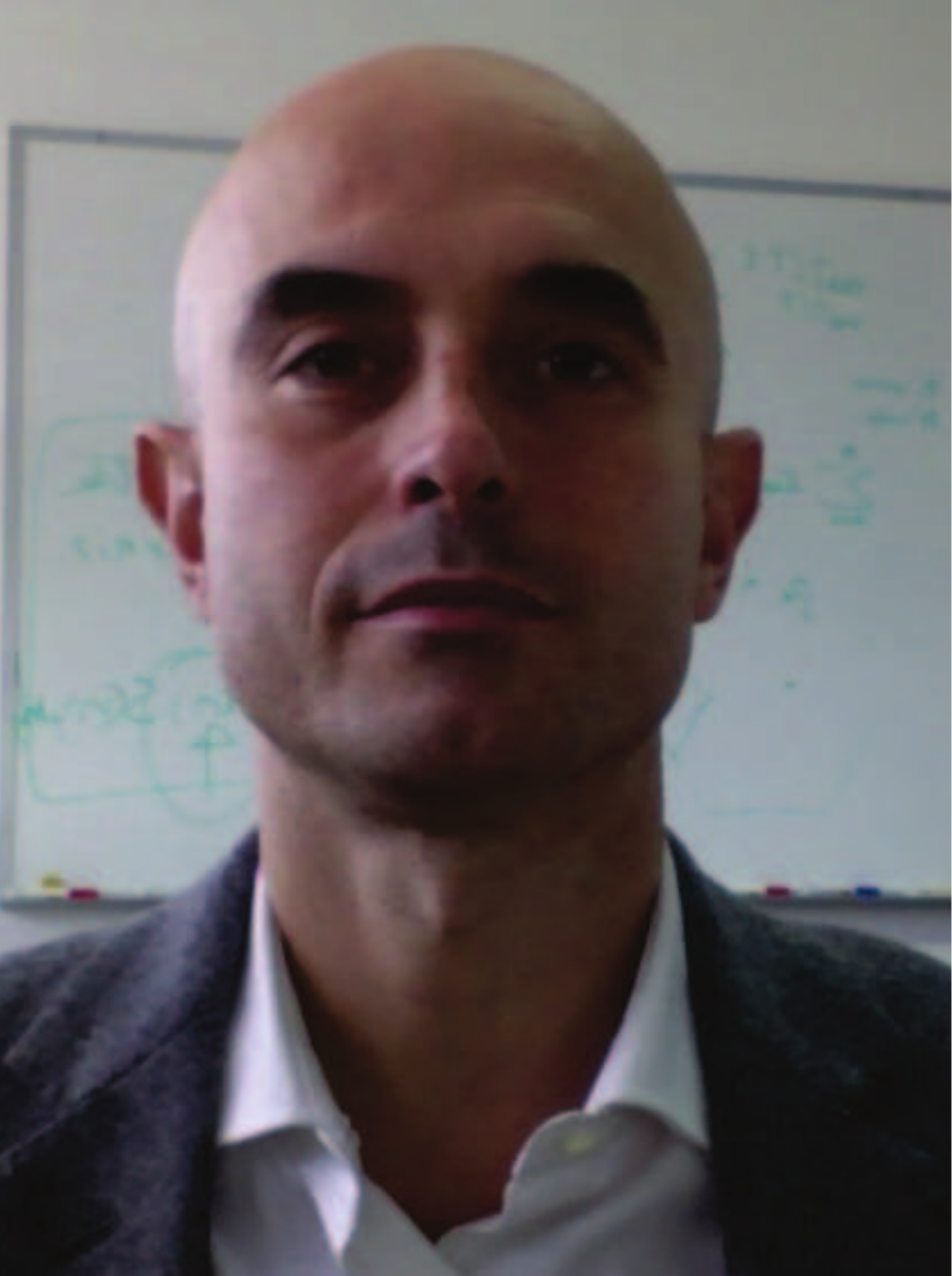}}]{Marco Moretti}(Member, IEEE)
received the degree in electronic engineering from the University of Florence, Florence, Italy, in 1995, and the Ph.D. degree from the Delft University of Technology, Delft, The Netherlands, in 2000. From 2000 to 2003, he was a Senior Researcher with Marconi Mobile. He is currently an Associate Professor with the University of Pisa, Pisa, Italy. His research interests include resource allocation for multicarrier systems, synchronization, and channel estimation. He is currently an Associate Editor of the IEEE TRANSACTIONS ON SIGNAL PROCESSING.
\end{IEEEbiography}

\begin{IEEEbiography}[{\includegraphics[width=1in,height=1.25in,clip,keepaspectratio]{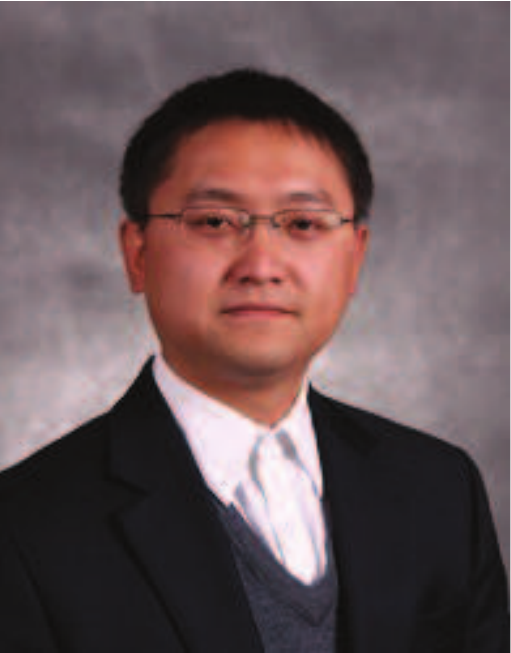}}]{Jian Xiong}(Member, IEEE)
received the B.Sc. and M.Sc. degrees from the University of Electronic Science and Technology of China, Chengdu, China, and the Ph.D degree from Shanghai Jiao Tong University (SJTU), Shanghai, China, in 1999, 2002, and 2006, respectively. He was a visiting scholar of Columbia University during 2015. He is currently an Associate Professor of the Image Communication and Networking Engineering Institute, SJTU. His current research interests include wireless wideband transmission technologies, networking, and caching technologies of converged wideband and broadcast systems. He has published over 70+ journal or conference papers and holds 40+ patents, including 25 awarded patents. He is the TPC Co-Chair of the IEEE International Symposium on Broadband Multimedia Systems and Broadcasting from 2010 to 2019.
\end{IEEEbiography}

\begin{IEEEbiography}[{\includegraphics[width=1in,height=1.25in,clip,keepaspectratio]{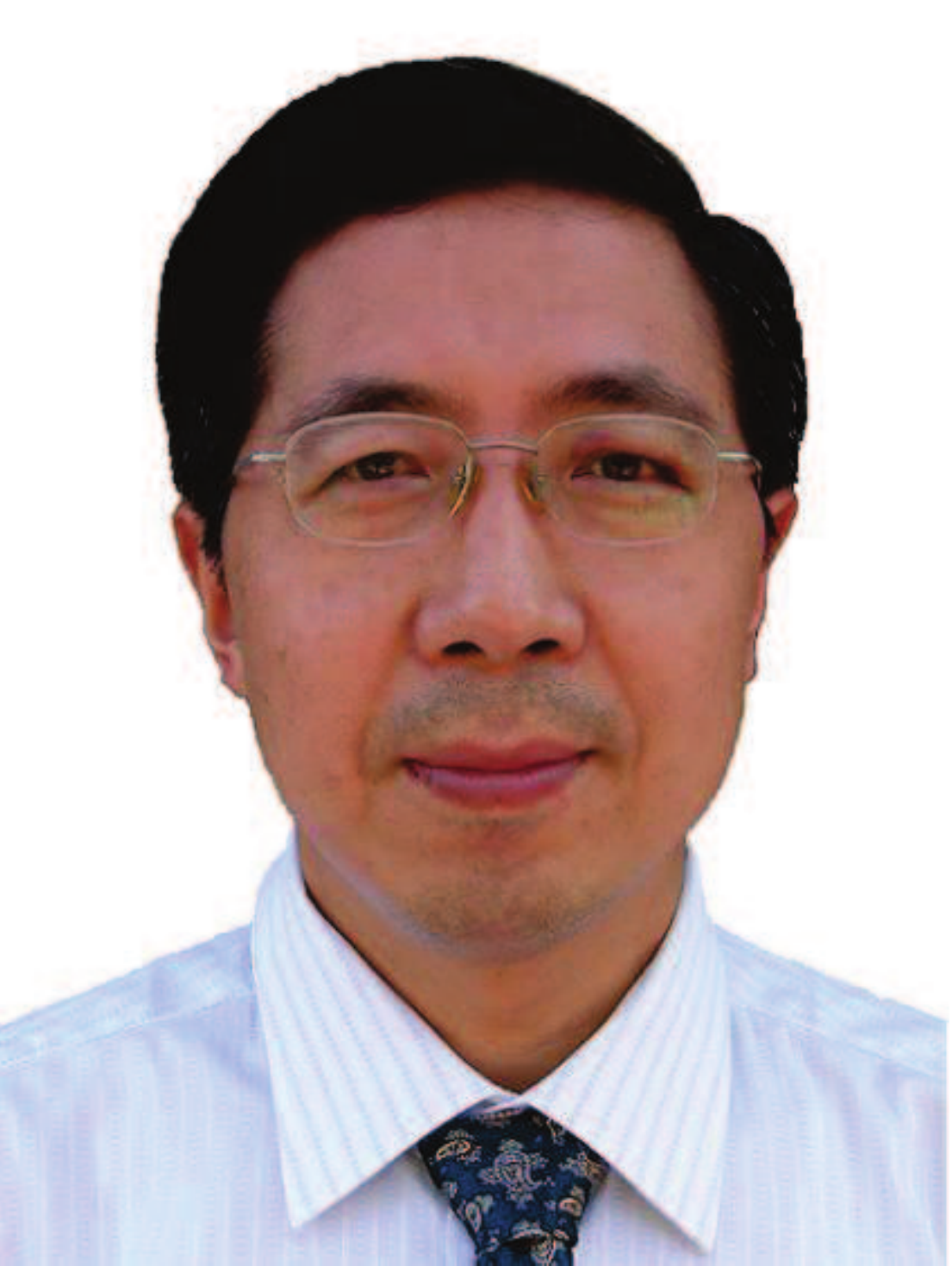}}]{Jiandong Li}(Fellow, IEEE)
received the B.E., M.S., and Ph.D. degrees in communications engineering from
Xidian University, Xi'an, China, in 1982, 1985, and 1991, respectively. He was a Visiting Professor with the Department of Electrical and Computer Engineering, Cornell University, from 2002 to 2003. He has been a faculty member of the School of Telecommunications Engineering, Xidian University, since 1985, where he is currently a Professor and the Vice Director of the Academic Committee, State Key Laboratory of Integrated Service Networks. His major research interests include wireless communication theory, cognitive radio, and signal processing. He was recognized as a Distinguished Young Researcher by NSFC and a Changjiang Scholar by the Ministry of Education, China, respectively. He served as the General Vice Chair of ChinaCom 2009 and the TPC Chair of the IEEE ICCC 2013.
\end{IEEEbiography}
\end{document}